\documentclass[10pt,a4paper]{article}

\usepackage{jheppub}

\usepackage{color}
\usepackage{colordvi}
\usepackage{changes}

\usepackage{epsfig,epsf}
\usepackage{amsmath}
\usepackage{amsthm}
\usepackage{amsfonts}
\usepackage{amssymb}
\usepackage{dsfont}
\usepackage{epstopdf}
\usepackage{multirow}

\usepackage{rotating}

\usepackage{marvosym}

\usepackage{todonotes}

\usepackage{subcaption}
\usepackage{array}   
\newcolumntype{C}{>{$}c<{$}}

\usepackage{slashed}

\usepackage[active]{srcltx}

\newcommand{\dhalf}{\frac{d}{2}}

\newcommand{\var}{\omega}

\newcommand \widebar [1] {\overline{#1}}

\def\II{\hbox{{1}\kern-.25em\hbox{l}}}

\DeclareMathOperator{\Li}{Li}

\newcommand \vev [1] {\langle{#1}\rangle}
\def\II{\hbox{{1}\kern-.25em\hbox{l}}}

\title{
\begin{flushright}
{\large \textnormal{DESY 20-116}}\\[2mm]
\end{flushright}
Two-loop coefficient function for DVCS: Vector contributions}

\author[a]{V. M. Braun,}
\author[b,c,a]{A. N. Manashov,}
\author[b]{S. Moch}
\author[a]{and J. Schoenleber}

\affiliation[a]{
   Institut f\"ur Theoretische Physik, Universit\"at
   Regensburg, D-93040 Regensburg, Germany}
\affiliation[b]{
   II.~Institut f\"ur Theoretische Physik, Universit\"at Hamburg,
   D-22761 Hamburg, Germany}
\affiliation[c]{
   St.Petersburg Department of Steklov
Mathematical Institute, 191023 St.Petersburg, Russia}

\emailAdd{vladimir.braun@ur.de}

\emailAdd{alexander.manashov@desy.de}

\emailAdd{sven-olaf.moch@desy.de}

\emailAdd{jakob.schoenleber@ur.de}

\abstract{
Using the approach based on conformal symmetry we calculate the two-loop coefficient function
for the vector flavor-nonsinglet contribution to deeply-virtual Compton scattering (DVCS).
The analytic expression for the coefficient function in momentum fraction space is presented in the
$\overline{\text{MS}}$ scheme.
The corresponding next-to-next-to-leading order correction to the Compton form factor $\mathcal{H}$ for a simple model
of the generalized parton distribution appears to be rather large: a factor two smaller than the next-to-leading order
correction, approximately $\sim 10$\% of the tree level result in the bulk of the kinematic range, for $Q^2=4$~GeV$^2$.
       }


\begin{document}
\maketitle

\section{Introduction}\label{sec:intro}

With the JLAB 12 GeV upgrade completed~\cite{Dudek:2012vr} and the Electron Ion Collider (EIC)~\cite{Accardi:2012qut} proposal
having received a major boost last year, there are bright perspectives for new generation of hadron physics studies in the coming
decade and beyond. The foreseen very high luminosity of these new machines will allow one to study hard exclusive and semi-inclusive reactions
with identified particles in the final state with unprecedented precision. Such processes are interesting as they allow one to
access hadron properties on a much more detailed level as compared to totally inclusive reactions. In this way one hopes to
understand the full three-dimensional proton structure by ``holographic imaging'' of quark and gluon distributions in distance and
momentum spaces.

Deeply-virtual Compton scattering (DVCS) \cite{Mueller:1998fv,Ji:1996nm,Radyushkin:1996nd} is generally accepted to be the
``gold-plated'' process that would have the highest potential impact for the transverse distance imaging. The general framework for
the QCD description of DVCS is based on collinear factorization in terms of generalized parton distributions (GPDs)
\cite{Diehl:2003ny,Belitsky:2005qn} and is well understood at the leading-twist level. The main challenge of these studies is that
the quantities of interest (GPDs) are functions of three variables (apart from the scale dependence). Their extraction requires
a massive amount of data and very high precision for both experimental and theory inputs. In the ideal case one would like to reach
the same level of accuracy as in inclusive reactions. The next-to-next-to leading order (NNLO)  analysis of parton distributions
and fragmentation functions has become the standard in this field~\cite{Accardi:2016ndt}, so that the NNLO precision for DVCS is
necessary as well.

The NNLO accuracy implies that one needs to derive three-loop evolution equations for GPDs and also calculate the
two-loop corrections to the coefficient functions (CFs) in the operator product expansion (OPE) of the DVCS amplitude.
The first part of this program, the three-loop evolution equation for flavor-nonsinglet GPDs, is completed
\cite{Braun:2017cih} and in this work we calculate the corresponding two-loop CF for
vector operators.
The two-loop axial vector CF
involves an additional subtlety of dealing with the $\gamma_5$ matrix
in off-forward reactions and will be considered in a separate publication.

From the theory point of view the main difference of DVCS from e.g. the classical case of the
deep-inelastic scattering (DIS) is that the target hadron has different momenta in the initial and the final state.
As a consequence, from the OPE point of view, one has to take into account additional contributions of operators containing
total derivatives. Such operators contribute both to operator mixing and to the CFs.
In conformal field theories the contributions of operators with total derivatives are related to the contributions
of the operators without total derivatives by symmetry transformations and do not need to be calculated
separately. Although conformal invariance of QCD is broken by quantum corrections, one can hope that the
symmetry of the Lagrangian can be used in some way in order to simplify the calculation.
The first (unsuccessful) attempt to predict the NLO evolution kernel from conformal symmetry \cite{Brodsky:1984xk}
was missing an important element: the scheme-dependent difference between the dilatation
and special conformal anomalies.
This problem was pointed out and solved by D.~M{\"u}ller~\cite{Mueller:1991gd}
who subsequently developed and applied (with collaborators) the conformal symmetry based technique to DVCS and GPDs.
In this way the complete two-loop mixing matrix was calculated for
twist-two operators in QCD~\cite{Mueller:1993hg,Mueller:1997ak,Belitsky:1997rh} and
the two-loop evolution kernels derived for GPDs~\cite{Belitsky:1998vj,Belitsky:1999hf,Belitsky:1998gc}.
It was also shown ~\cite{Belitsky:1997rh} that conformal symmetry allows one to obtain
the one-loop CF in DVCS from the known CF in DIS.
%
%
 Thus the complete NLO calculation of the DVCS amplitudes to the leading-twist accuracy is available. An estimate of the NNLO
contribution in a special renormalization scheme (``conformal scheme'') was also
attempted~\cite{Mueller:2005nz,Kumericki:2006xx,Kumericki:2007sa}.

In Ref.~\cite{Braun:2013tva} we suggested a somewhat different implementation of the same
idea. Instead of studying conformal symmetry {\it breaking} in the physical
theory~\cite{Mueller:1993hg,Mueller:1997ak,Belitsky:1997rh} we proposed to make use of the
{\it exact} conformal symmetry of large-$n_f$ QCD in $d=4-2\epsilon$ dimensions
at the Wilson-Fischer fixed point at critical coupling~\cite{Braun:2018mxm}.
Due to specifics of the minimal subtraction scheme ($\overline{\text{MS}}$)
the renormalization group equations (RGEs) in
the physical four-dimensional theory inherit conformal symmetry of the $d=4-2\epsilon$ theory
so that the evolution kernel commutes with the generators of conformal transformations.
This symmetry is exact, however, the generators are modified by quantum corrections and differ from their canonical form.
The consistency relations that follow from the conformal algebra can be used in order to restore  the
$\ell$-loop off-forward kernel from the $\ell$-loop anomalous dimensions and the $(\ell-1)$-loop
result for the deformation of the generators, which is equivalent to the statement in
Ref.~\cite{Mueller:1991gd}. The two-loop expression for the generator of special conformal transformations
for flavor-nonsinglet operators is obtained in~\cite{Braun:2016qlg} and is the main ingredient in the
calculation of the three-loop evolution kernel in Ref.~\cite{Braun:2017cih}.
For the flavor-singlet case one had first to clarify the structure of the contributions
of gauge non-invariant operators to conformal Ward identities, see Ref.~\cite{Braun:2018mxm}.
We used this result in~\cite{Braun:2019qtp} to derive the two-loop flavor-singlet
evolution equation for light-ray operators and confirmed in this way the expression for the evolution kernel
obtained originally in~\cite{Belitsky:1999hf}.

This work is devoted to the conformal symmetry based approach to the calculation of the DVCS CFs.
The presentation is organized as follows.
Sect.~\ref{sec:general} is introductory, it contains
general definitions, our notation and conventions.
In Sect.~\ref{sec:framework} we present the general framework for the calculation
of CFs in the OPE of two electromagnetic currents using conformal
symmetry of QCD at the Wilson-Fischer fixed point in non-integer dimensions.
The main statement is that calculation of the  $\ell$-loop off-forward CF
can be reduced to the  $\ell$-loop forward CF, known from DIS, and
the $(\ell-1)$-loop calculation of the off-forward CF in $4-2\epsilon$ dimensions,
including terms $\mathcal{O}(\epsilon^{\ell-1})$.
The one-loop example for the application of this machinery is considered in Sect.~\ref{sec:one-loop}
where we reproduce the corresponding well-known expression~\cite{Ji:1997nk}.
The following Sect.~\ref{sec:two-loop} contains our main result: the derivation of the two-loop
CF for vector operators. The analytic expression for the CF is presented in momentum fraction space in the
$\overline{\text{MS}}$ scheme.
We find that the CF in the conformal scheme satisfies the reciprocity relation
\cite{Dokshitzer:2005bf,Basso:2006nk,Alday:2015eya,Alday:2015ewa} that arises in the sum
of many contributions and provides one with a highly nontrivial check of the results,
in particular for the two-loop conformal anomaly~\cite{Braun:2016qlg}.
Numerical estimates of the size of the NNLO correction for two popular GPD models are
presented in  Sect.~\ref{sec:numerics}. We conclude in Sect.~\ref{sec:summary} with a short summary and outlook.
Some useful integrals are collected in the Appendix.

\section{DVCS kinematics, notation and conventions}\label{sec:general}
The amplitude of the DVCS process is given by a matrix element of the time-ordered product of two electromagnetic currents
\begin{align}
\mathcal{A}_{\mu\nu}(q,q',p) &= i\int d^4 x\, e^{-iqx} \vev{p'|T\{j^{\rm em}_\mu(x)j^{\rm em}_\nu(0)\}|p}\,.
\label{DVCS}
\end{align}
Here $q,q'$ are the momenta of the virtual (incoming)  and real (outgoing) photons and $p,p'$  are the target (nucleon)
momenta in initial and final  states.  We use the photon momenta to define the longitudinal plane spanned by two light-like
vectors~\cite{Braun:2012bg},
\begin{align}
n=q', && \bar n =-q+(1-\tau)q',
\end{align}
where $\tau=t/(Q^2+t)$, $Q^2=-q^2$. In the leading-twist approximation the DVCS amplitude  can be written
as a sum of vector and axial-vector contributions~\footnote{For the complete expression see~\cite{Braun:2012bg}.}
\begin{align}
\mathcal{A}_{\mu\nu}= -g_{\mu \nu}^\perp  V  + \epsilon_{\mu\nu}^\perp  A +\ldots
\label{def:V}
\end{align}
where
\begin{align}\label{structures}
g_{\mu\nu}^\perp=g_{\mu\nu}-\frac{q_\mu q'_\nu+q'_\mu q_\nu}{(qq')}+{q'_\mu}q'_\nu\frac{q^2}{(qq')^2}\,,
&&
\epsilon_{\mu\nu}^\perp=\frac1{(qq')}\epsilon_{\mu\nu\alpha\beta}{q^\alpha q'^\beta}\,.
\end{align}
The ellipses stand for the higher twist contributions of helicity-flip amplitudes and terms $\sim q^\prime_\mu$, which do not
contribute to physics observables  thanks to electromagnetic Ward identities.

In this work we will consider flavor-nonsinglet contributions to the vector amplitude $V$ in the leading-twist approximation.
To this accuracy it can be written as a convolution of the CF with the corresponding GPD
\begin{align}\label{C-DVCS}
V(\xi,Q^2)=\sum_q e_q^2 \int_{-1}^1 \frac{dx}{\xi} C(x/\xi,Q^2/\mu^2)\, F_q(x,\xi,t,\mu)\,.
\end{align}
Here and in what follows $q$ is a quark flavor, $q=u,d,s,\ldots$,
$\mu$ stands for the factorization scale in the $\overline{\rm MS}$ scheme and
we use established conventions for the kinematical variables
\begin{align}
\xi=-\Delta_+/2P_+, && t=\Delta^2, &&\Delta=p'-p,&& P=(p+p')/2,&& a_+\equiv a_\mu n^\mu\,.
\end{align}
 The GPD $F_q$ is defined by the appropriate matrix element of the light-ray quark-antiquark operator,
\begin{align}\label{l-ray-operator}
\mathcal O_q(z_1 n,z_2n)=\bar q(z_1 n)\slashed{n}[z_1n,z_2n]q(z_2n)\,,
\end{align}
where $[z_1n,z_2n]$ is a Wilson line. The matrix element can, in general, be parameterized as
\begin{align}\label{H-def}
\langle p'|\mathcal O_q(z_1,z_2)|p\rangle =2P_+\int_{-1}^1 dx\,  e^{-iP_+\xi(z_1+z_2) +i P_+x (z_1-z_2)} \, F_q(x,\xi)\,.
\end{align}
The expression for $F_q$ depends on the spin of the target (e.g. nucleon vs. pion or ${}^4$He nucleus) whereas the CF
is the same in all cases. For the most important case of DVCS off the nucleon, $F_q$
can further be decomposed in contributions of the standard GPDs, $H$, $E$, as follows~\cite{Diehl:2003ny}
\begin{align}
F_q(x,\xi)=\frac1{2P_+}\left[H_q(x,\xi,t) \bar u(p')\gamma_+u(p)+ E_q(x,\xi,t)
\bar u(p')\frac{i\sigma^{+\nu}\Delta_\nu}{2m_N}u(p)\right].
\end{align}
Note that only the charge conjugation even $C=+1$ part of the GPDs $F_q^{(+)}(x,\xi,t) = F_q(x,\xi,t)- F_q(-x,\xi,t)$ can contribute
to the vector amplitude, which is reflected in antisymmetry of the CF
\begin{align}
    C(- x/\xi,Q^2/\mu^2) = -  C(x/\xi,Q^2/\mu^2)\,.
\end{align}
The CF can be calculated in perturbation theory
\begin{align}
   C(x/\xi,Q^2/\mu^2) &=  C^{(0)}(x/\xi) + a_s  C^{(1)}(x/\xi,Q^2/\mu^2) + a_s^2  C^{(2)}(x/\xi,Q^2/\mu^2) +\ldots
   \, ,
\label{eq:series}
\end{align}
where
\begin{align}
  a_s &= \frac{\alpha_s(\mu)}{4\pi}\,,
\end{align}
and the first two terms in this series are well known (see e.g.~\cite{Belitsky:2005qn})
\begin{align}
   C^{(0)}(x/\xi)  &= \frac{\xi}{\xi-x} - \frac{\xi}{\xi+x}\,,
\notag\\[2mm]
C^{(1)}(x/\xi,Q^2/\mu^2) &= \frac{2C_F \xi}{\xi-x}
\biggl[
\ln \frac{Q^2}{2\mu^2}\left(\frac32+\ln\left(\frac{\xi-x}{2\xi}\right)\right)
   \notag\\
   &\quad - \frac92 -\frac12 \ln^2 2
    +\left[\frac12
    \ln\left(1-\frac x{\xi}\right)
    - \frac32 \frac{\xi-x}{\xi+x} \right]\ln \left(1-\frac x{\xi}\right)
\biggr]-(x\leftrightarrow -x)\,.
\label{CF-oneloop}
\end{align}
In this work we calculate the two-loop expression $C^{(2)}(x/\xi,Q^2/\mu^2)$.

It is very important that the CFs only depend on the ratio $x/\xi$ and are real functions at $|x/\xi| <1$ (ERBL region),
and can be continued analytically to the DGLAP region  $|x/\xi| \ge 1$ using the $\xi \to \xi-i\epsilon$ prescription.
This property holds to all orders and allows one to simplify the calculation assuming $\xi=1$ in which case the CF in DVCS coincides
(after a redefinition of kinematic variables) with
the CF in the transition form factors $\gamma^\ast\gamma \to\,\text{meson}$ with appropriate quantum numbers.
If the latter is known, the corresponding CF in DVCS is obtained with the substitution $x\to x/(\xi-i\epsilon)$.

\section{General framework}\label{sec:framework}

The OPE for the product of currents has a generic form, schematically
\begin{align}
T\{j^{\rm em}_\mu(x)j^{\rm em}_\nu(0)\} &= \sum_{N,k} C_{N,k} \partial_+^k\mathcal O_N(0)
\end{align}
where $\mathcal O_N(0)$ are local operators of increasing dimension and $C_{Nk}$ are the corresponding CFs. For forward matrix
elements, like in DIS, contributions of the operators with total derivatives vanish identically and can be omitted. Thus only the
sum over $N$ remains and the necessary CFs  $C_{N} \equiv C_{N,k=0}$ are known to three-loop accuracy \cite{Vermaseren:2005qc}. In
off-forward reactions, like DVCS, operators with total derivatives have to be included and one needs to calculate their
coefficients, $C_{N,k}$ with $k\ne 0$, as well. In conformal field theories a direct calculation is not needed since the CFs of
operators with total derivatives are related to the CFs of the operators without total derivatives by the symmetry transformations
(for the special choice of the operator basis $\mathcal{O}_N$). We will show how to restore $C_{N,k}$ from $C_{N,0}$ using
conformal algebra in what follows.

QCD in four space-time dimensions, however, is not a conformal theory. The idea is to consider QCD in non-integer $d=4-2\epsilon$
dimensions at the intermediate step, for the specially chosen (critical) value of the coupling $\alpha_s^\ast$ such that the
$\beta(\alpha_s^*)=0$ (Wilson-Fisher fixed point). This theory is conformally invariant~\cite{Braun:2018mxm} and all
renormalization constants/anomalous dimensions for composite operators in this theory in minimal subtraction schemes coincide with
the renormalization constants/anomalous dimensions of the corresponding operators for the ``real'' QCD in
$d=4$~\cite{Braun:2013tva}.

One can consider, formally, the DVCS process in a generic $4-2\epsilon$-dimensional theory. All definitions in
Sec.~\ref{sec:general} can be taken over without modifications except for that the CFs acquire an $\epsilon$-dependence so that
     $$C(x/\xi, Q^2/\mu^2,a_s)\mapsto C(x/\xi, Q^2/\mu^2,a_s,\epsilon)\,,$$
and their perturbative expansion involves $\epsilon$-dependent coefficients:
\begin{align}
C(a_s,\epsilon)& =C_0 + a_s\, C^{(1)}(\epsilon) +  a_s^2\, C^{(2)}(\epsilon)+ \mathcal{O}(a_s^3)\,,
\notag\\
C^{(k)}(\epsilon)& = C^{(k)} + \epsilon\, C^{(k,1)} + \epsilon^2 C^{(k,2)} + \mathcal{O}(\epsilon^3)\,.
\label{eq:generic-d}
\end{align}
Note that the tree-level CF $C^{(0)}$ does not depend on $\epsilon$.

We are interested in the CF in four dimensions \eqref{eq:series} as a function of the coupling,
whereas methods of conformal field theories allow one
to calculate $C_\ast = C(\alpha_s^*,\epsilon)$ on the line in the $(\epsilon, \alpha_s)$ plane
where $\beta(\alpha_s^\ast)=0$ so that $\alpha_s^\ast =\alpha_s^\ast(\epsilon)$ or, equivalently,
\begin{align}
\epsilon_\ast &= \epsilon(a_s^\ast) = -\big(\beta_0 a_s^\ast + \beta_1(a_s^\ast)^2 + \ldots\big)\,, \qquad
\beta_0 = \frac{11}{3}N_c -\frac23 n_f\,,
\end{align}
with $N_c$ and $n_f$ the numbers of colors and light flavors, respectively.
Trading the $\epsilon$-dependence for the $a_s^\ast$ dependence one can write the
CF at the critical point as an expansion in the  coupling alone,
\begin{align}
C_\ast(a_s) & = C(a_s,\epsilon_\ast)= C^{(0)} +a_s C_\ast^{(1)} +a_s^2 C_\ast^{(2)}+\mathcal{O}(a_s^3)\,,
\label{eq:critical-expand}
\end{align}
where, obviously,
\begin{align}
   C_\ast^{(1)} &= C^{(1)}\,,
\notag\\
   C_\ast^{(2)}  &= C^{(2)} - \beta_0 C^{(1,1)}\,.
\label{eq:d=4-restored1}
\end{align}
The coefficients  $C_\ast^{(k)}$ can be related to the known CFs for DIS (not without effort) thanks to conformal invariance.
Thus
\begin{align}
 C^{(1)} & = C_\ast^{(1)}\,,
\notag\\
 C^{(2)} & = C_\ast^{(2)} +\beta_0 C^{(1,1)}\,.
\label{eq:d=4-restored}
\end{align}
In other words, the one-loop CF in $d=4$ coincides with the one-loop CF in conformal QCD in $4-2\epsilon$ dimensions and in
order to find the two-loop CF in $d=4$ one needs to know the corresponding result in conformal QCD and, in addition, terms of order
$\epsilon$ in the one-loop CF. Since all necessary one-loop integrals can be calculated in terms of $\Gamma$-functions for
arbitrary space-time dimensions, the latter calculation is rather straightforward and in what follows we will present the final
result only. The calculation of $C^{(2)}_\ast$ presents the main challenge and will be discussed in detail.

The expansion in Eq.~\eqref{eq:d=4-restored1} or \eqref{eq:d=4-restored} can obviously be continued to higher orders.
The general statement is that the $\ell$-loop
off-forward CF in QCD in $d=4$ in the $\overline{\text{MS}}$ scheme can be obtained from the corresponding result in conformal
theory (alias from the CF in the forward limit), adding terms proportional to the QCD beta-function. Such extra terms require the
calculation of the corresponding $\ell-1$-loop off-forward CF in $d=4-2\epsilon$ dimensions to the $\sim
\mathcal{O}(\epsilon^{\ell-1})$ accuracy.

\subsection{Conformal OPE}\label{sect:COPE}

Retaining the contributions of vector operators only,
the most general expression for the OPE of the product of two electromagnetic currents
to the twist-two accuracy has the form
\begin{align}\label{COPE}
\text{T}\,\big\{j^\mu(x_1) j^\nu(x_2)\big\} & =
\sum_{N,\text{even}}\frac{\mu^{\gamma_N}}{(-x_{12}^{2}+i0)^{t_N}}
 \int_0^1 du
 \Biggl\{ -\frac12 A_N(u) \left( g^{\mu\nu} - \frac{2x_{12}^\mu x_{12}^\nu}{x_{12}^2}
 \right) + B_N(u) g^{\mu\nu}
 \notag\\
&\quad
 +C_N(u) x_{12}^\nu\partial_1^\mu - C_N(\bar u) x_{12}^\mu
 \partial_2^\nu
+D_N(u)x_{12}^2 \partial_1^\mu\partial_2^\nu \Biggr\}
\mathcal O_N^{x_{12}\ldots x_{12}}(x_{21}^u)\,,
\end{align}
where
\begin{align}
  \partial^\mu_k = \frac{\partial}{\partial x^\mu_k}, && x_{12} = x_1-x_2\,, && \bar u = 1-u\,, && x_{21}^u=\bar u x_2+ u x_1\,,
\end{align}
and
\begin{align}
 \mathcal O_N^{x\ldots x}(y) = x_{\mu_1}\ldots x_{\mu_N}  \mathcal O_N^{\mu_1\ldots\mu_N}(y)\,,
\end{align}
where $\mathcal O_N^{\mu_1\ldots\mu_N}(y)$ are the leading-twist conformal operators
that transform in the proper way under conformal transformations
\begin{align}
   [\mathbb{K}_\mu, \mathcal O_N^{x\ldots x}(y)] =
   \left( 2y_\mu  y^\nu \frac{\partial}{\partial y^\nu}-y^2 \frac{\partial}{\partial y^\mu}
   +2\Delta_N y_\mu  +2y^\nu\left(x_\mu \frac{\partial}{\partial x^\nu} -
   x_\nu \frac{\partial}{\partial x^\mu}\right)\right) \mathcal O_N^{x\ldots x}(y).
\end{align}
Here and below $N$ is the operator spin, $\Delta_N$ is its scaling dimension,
$\Delta_N=d_\ast+N-2+\gamma_N$ where $d_\ast = 4-2\epsilon_\ast$,
$\gamma_N=\gamma_N (a_s)$ is the anomalous dimension,
$t_N= 2 -\epsilon_\ast -\frac12\gamma_N(a_s)$ is the twist and
$j_N = N + 1 -\epsilon_\ast +\frac12\gamma_N(a_s)$ is the conformal spin.
We have separated in Eq.~\eqref{COPE} the scale factor $\mu^{\gamma_N}$ to make the invariant functions
$A_N(u),\ldots, D_N(u)$  dimensionless.
Note that only vector operators with even spin $N$ contribute to the expansion.

The conditions of conformal invariance and current conservation $\partial^\mu j_\mu=0$ lead to constraints on the functional form
and also certain relations between the invariant functions $A_N(u),\ldots, D_N(u)$ in Eq.~\eqref{COPE}. One obtains
\begin{align}
A_N(u) = a_N\, u^{j_N-1}\bar u^{j_N-1}\, , && B_N(u) = b_N\, u^{j_N-1} \bar u^{j_N-1}\, .
\label{AB}
\end{align}
The remaining functions $C_N(u)$  and $D_N(u)$ are given by
the following expressions:
\begin{align}
C_N(u) & = u^{N-1}\int_u^1\frac{dv}{v^N}v^{j_N} \bar v^{j_N-2} \left(c_N -\frac{b_N}v\right),
 \notag\\
D_N(u)  &= -\frac1{N-1} \int_0^1{dv}( v  \bar v)^{j_N-1}\left[\theta(v-u) \left(\frac{u} v\right)^{N-1}%
 + \theta(\bar v-\bar u) \left(\frac{\bar u}{\bar v}\right)^{N-1} %
 \right]\left(d_N -\frac{c_N-b_N}{2 v \bar v} \right).
\end{align}
The coefficients $c_N$ and $d_N$ are not independent and are given in terms of $a_N$ and  $b_N$ by linear relations
\begin{align}
\left(j_N-1\right)\, a_N & = 2\,t_N\left(c_N -b_N\right)\, ,
\notag\\
2\,(t_N-1)\, d_N &= -\frac12 a_N (N-j_N)-\gamma_N b_N + \big(j_N-2+2t_N\big) (c_N -b_N)\, .
\end{align}
Thus the form of the OPE of the product of two conserved spin-one currents in a
conformal theory is fixed up to two constants, $a_N(a_s)$ and $b_N(a_s)$, for each (even) spin $N$.
In QCD the expansion  of  $a_N(a_s) $ starts at order $\mathcal{O}(a_s)$, so that
also  $c_N-b_N= \mathcal{O}(a_s)$ and $d_N=\mathcal{O}(a_s)$.

\subsection{Relating DIS and DVCS}

It is convenient to fix the normalization of the leading-twist conformal operators such that
\begin{align}
   \mathcal O_N^{\mu_1\ldots\mu_N}(0) = i^{N-1}\bar q(0) \gamma^{\{\mu_1} D^{\mu_2}\ldots D^{\mu_N\}} q(0)
   +\,\text{total~derivatives}\,,
\label{eq:normalization}
\end{align}
where $D^{\mu} = \partial^{\mu} + i g A^{\mu}$ and
$\{\ldots\}$ denotes the symmetrization of all enclosed Lorentz indices and the subtraction
of traces. In this way the forward matrix elements of these operators are related to moments of
quark parton distributions (PDFs)
\begin{align}
   \langle p| \mathcal O_N^{\mu_1\ldots\mu_N}(0)|p\rangle = p^{\{\mu_1}\ldots p^{\mu_N\}} f_N \,.
\end{align}
Using this parametrization and taking the Fourier transform of the forward matrix element of Eq.~\eqref{COPE} one obtains
the OPE for the forward Compton tensor in a generic conformal theory
\begin{flalign}
T_{\mu\nu}(p,q) &\equiv i\int d^dx\, e^{-iqx}\langle p|T(j_\mu(x)j_\nu(0)|p\rangle
\notag\\
&=
\sum_{N,\text{even}}\!\!
f_N\!\left(\frac{2pq}{Q^2}\right)^N\!\!\!
\left(\frac{\mu}{Q}\right)^{\gamma_N}
\!\biggl[
\left(-g_{\mu\nu} + \frac{q_\mu q_\nu}{q^2}\right)c_{1N}(a_*)
+\frac{(q_\mu\!+\!2x_Bp_\mu)(q_\nu\!+\!2x_Bp_\nu)}{Q^2} c_{2N}(a_*)
\biggr],
\end{flalign}
where $x_B = Q^2/(2qp)$ is the Bjorken scaling variable and
\begin{align}
 c_{1N} &= i^N\pi^{d/2} 2^{\gamma_N} B(j_N,j_N) \frac{\Gamma(N+\gamma_N/2)}{\Gamma(t_N)}
\left( \frac{t_N-1}{2t_N}a_N-b_N \right),
\notag\\
c_{2N} & = i^N\pi^{d/2} 2^{\gamma_N} B(j_N,j_N) \frac{\Gamma(N+\gamma_N/2)}{\Gamma(t_N)}
\left(-b_N + \frac{2N+d -t_N-1}{2t_N} a_N\right).
\end{align}
Here and below  $B(j_N,j_N)$ is the Euler Beta function.

The same expansion in QCD is usually written as~\cite{Vermaseren:2005qc}
\begin{align}
 T_{\mu\nu}(p,q) &=
\sum_{N,\text{even}}\!\!f_N\!\left(\frac{2pq}{Q^2}\right)^N
\biggl[\left(g_{\mu\nu} - \frac{q_\mu q_\nu}{q^2}\right)C_{L}\left(N,\frac{Q^2}{\mu^2},a_s\right)
\notag\\&\quad
-\left(g_{\mu\nu} - p_\mu p_\nu \frac{4x_B^2}{Q^2} - (p_\mu q_\nu + p_\nu q_\mu)\frac{2x_B}{Q^2}\right)
C_{2}\left(N,\frac{Q^2}{\mu^2},a_s\right)\biggr],
\end{align}
(we drop electromagnetic charges and the sum over flavors), so that we can identify
\begin{align}
c_{2N}(a_s)\left(\frac{\mu}{Q}\right)^{\gamma_N}  & = C_{2}\left(N,\frac{Q^2}{\mu^2},a_s,\epsilon_\ast\right),
\notag\\
c_{1N}(a_s)\left(\frac{\mu}{Q}\right)^{\gamma_N}  & = C_{2}\left(N,\frac{Q^2}{\mu^2}, a_s,\epsilon_\ast\right)-
C_{L}\left(N,\frac{Q^2}{\mu^2},a_s,\epsilon_\ast\right)
\equiv C_{1}\left(N,\frac{Q^2}{\mu^2},a_s,\epsilon_\ast\right).
\end{align}
The coefficient functions $C_2$ and $C_L$ contribute to the OPE for the structure functions $F_2$ and $F_L$, respectively, and are
known to third order in the coupling.

Next, let us consider the DVCS process~\eqref{DVCS}.
In comparison to forward scattering there are two modifications.
First, the position of the operator $\mathcal O_N(ux)$ in Eq.~\eqref{COPE} (we assume here $x_1\mapsto x$, $x_2\mapsto 0$)
becomes relevant since
\begin{align}
 \langle p'|\mathcal O_N(ux)|p\rangle= e^{-i u (x\cdot (p-p'))}\langle p'|\mathcal O_N(0)|p\rangle
 = e^{i u (x\cdot \Delta)}\langle p'|\mathcal O_N(0)|p\rangle
\, ,
\end{align}
and effectively results in a shift of the momentum in the Fourier integral $q\mapsto q-u\Delta$. Second, the matrix element becomes
more complicated and can be parameterized as
\begin{align}
\langle p'|n^{\mu_1} \ldots n^{\mu_N}
\mathcal O_{\mu_1\ldots\mu_N}(0)|p\rangle & =\sum_k \left(-\frac12\right)^k f_N^{(k)} P_+^{N-k} \Delta_+^k =  P_+^N f_N(\xi)\,,
\notag\\
f_N(\xi) &\equiv \sum_k f_N^{(k)}\xi^k\,, \qquad f_N^{(0)}= f_N^{\rm DIS}\,.
\label{fN(xi)}
\end{align}
Hence one needs a more general Fourier integral
\begin{eqnarray}
\lefteqn{\hspace*{-2cm}
 \sum_{k=0}^N \left(-\frac{1}{2}\right)^k f_N^{(k)} \int d^d x\,e^{-i(q-u\Delta)x}
 \frac{ \Gamma[t_N]}{[-x^2+i\epsilon]^{t_N}} (x\cdot P)^{N-k}(x\cdot\Delta)^k  =}
\nonumber\\
 &&\hspace*{2cm}{}~=~
i^{N-1} 2^{\gamma_N} \pi^\dhalf \frac{\Gamma[\tfrac12 \gamma_N + N]}{\bar u^{N+\frac12\gamma_N}Q^{\gamma_N}}
\left(\frac{2Pq}{Q^2}\right)^N
 \sum_{k=0}^N f_N^{(k)}\xi^k\, ,
\end{eqnarray}
where we neglected all power-suppressed corrections $\Delta^2/Q^2$ and also used that to this accuracy $(q-u\Delta)^2 = -\bar u Q^2$.

Up to the obvious replacement $p\mapsto P$ there are two differences to the forward case (DIS): an extra factor $\bar
u^{-N-\frac12\gamma_N}$, and the matrix element $f_N \to  f_N(\xi)$. Note that the leading-twist DVCS vector amplitude
\eqref{def:V} corresponds to the Lorentz structure $g^\perp_{\mu\nu}$ in Eq.~\eqref{structures} in the Compton tensor \eqref{DVCS}, and
can be traced by contributions $\sim g_{\mu\nu}$ (in momentum space). Starting from the position space expression
in Eq.~\eqref{COPE},  such terms can only originate from structures $\sim g_{\mu\nu}$ and $\sim x_\mu x_\nu$ which involve the invariant
functions $A_N(u)$ and $B_N(u)$ with the same $u$-dependence \eqref{AB}. The extra factor $\bar u^{-N-\frac12\gamma_N}$, therefore,
results in both cases in the following modification:
\begin{align}
B(j_N,j_N) = \int du\, (u\bar u)^{j_N-1}\mapsto \int du (u\bar u)^{j_N-1} \bar u^{-N - \frac12\gamma_N}
= B(j_N,\tfrac{d}{2}-1)\,,
\end{align}
where (see above) $j_N=N+ 1-\epsilon_\ast+\frac12\gamma_N$.
Since this modification affects the contribution of the structures $A_N(u)$ and $B_N(u)$ in the same way, we actually do not
need to consider them separately.
Thus the OPE for the DVCS  amplitude $V$ in conformal QCD in non-integer dimensions can be written as
\begin{align}\label{V-COPE}
V(\xi,Q^2) &=%
\sum_{N,{\rm even}}%
f_N(\xi)\left(\frac{2Pq}{Q^2}\right)^N C_{1}\left(N,\frac{Q^2}{\mu^2},a_s,\epsilon_\ast\right)\,
\frac{\Gamma(\dhalf-1)\Gamma(2j_N)}{\Gamma(j_N)\Gamma(j_N+\dhalf-1)}
\notag\\
&=%
\sum_{N,{\rm even}}%
f_N(\xi)\left(\frac{1}{2\xi}\right)^N C_{1}\left(N,\frac{Q^2}{\mu^2},a_s,\epsilon_\ast\right)\,
\frac{\Gamma(\dhalf-1)\Gamma(2j_N)}{\Gamma(j_N)\Gamma(j_N+\dhalf-1)}
\end{align}
and is completely determined by the forward-scattering coefficients $C_{1}(N)$ (in $d=4-2\epsilon_\ast$). For $d=4$ the above
expression agrees with \cite[Eq.~(22)]{Belitsky:1997rh}.
 In the next section
we show how the DVCS coefficient function in momentum fraction space~\eqref{C-DVCS} can be obtained starting from this
representation.

 \subsection{Coefficient function in momentum fraction space}

GPDs are defined as matrix elements \eqref{H-def} of nonlocal light-ray operators \eqref{l-ray-operator} so that
in order to relate the CFs in position or momentum fraction space to the CFs in the OPE one needs an expansion of the type
\begin{align}
[\mathcal O(z_1,z_2)]=\sum_{Nk} \Psi_{Nk}(z_1,z_2)\,\partial_+^k [\mathcal O_N(0)]\,.
\label{LROPE}
\end{align}
Here $[\ldots]$  stands for renormalization in the $\overline{\rm MS}$ scheme in $d=4-2\epsilon_\ast$ and
$\mathcal O_N \equiv n_{\mu_1}\ldots  n_{\mu_N} \mathcal O_N^{\mu_1\ldots\mu_N}$ are conformal operators that satisfy the RGE
\begin{align}
\Big(\mu\partial_\mu +\gamma_N(a_s)\Big)[\mathcal O_N]&=0\,.
\label{eq:RGEloc}
\end{align}
We will tacitly assume that conformal operators are normalized as in Eq.~\eqref{eq:normalization}.
The light-ray operator $[\mathcal O(z_1,z_2)]$, in turn, satisfies the RGE of the form
\begin{align}
\Big(\mu\partial_\mu +\mathbb H(a_s)\Big)[\mathcal O(z_1,z_2)]=0\,,
\end{align}
where $\mathbb H$ (evolution kernel) is an integral operator acting on the coordinates $z_1,z_2$.
Conformal symmetry ensures that $\mathbb H$ commutes
with the generators of $SL(2,\mathbb R)$ subgroup of the conformal group.
Translation-invariant polynomials $z_{12}^N \equiv (z_1-z_2)^N$ are eigenfunctions of the
evolution kernel and the corresponding eigenvalues define the anomalous dimensions
\begin{align}
  \mathbb H(a_s) z_{12}^{N-1} = \gamma_N(a_s) z_{12}^{N-1}\,.
\end{align}

The expansion coefficients $\Psi_{Nk}$ in Eq.~\eqref{LROPE} are homogeneous polynomials
in $z_1,z_2$ of degree $N+k-1$ and are given by a repeated application of the
generator of special conformal transformations $S_+$ to the coefficient of the conformal
operator, $\Psi_{N,k} (z_1,z_2) \sim S^k_+ z_{12}^N$. The problem is that $S_+$ in the
interacting theory in the $\overline{\rm MS}$  scheme contains a rather complicated
conformal anomaly term $z_{12}\Delta_+$~\cite{Braun:2016qlg} so that finding explicit
expression for $S^k_+ z_{12}^N$ is difficult.

The way out is to do a rotation to the ``conformal scheme'' at the intermediate step using a similarity transformation
defined in \cite{Braun:2017cih}
\begin{align}
[\mathbf{O}(z_1,z_2)]= \mathrm{U}\, [\mathcal{O}(z_1,z_2)]\,, && \mathbb H =
\mathrm{U}^{-1} \mathbf{H} \mathrm{U}\,, && S_{\pm,0} = \mathrm{U}^{-1} \mathbf{S}_{\pm,0} \mathrm{U}\,.
\end{align}
Note that $\mathbb{H}$ and $\mathbf{H}$ obviously have the same eigenvalues (anomalous dimensions).
Going over to the ``boldface'' operators can be thought of as a change of the
renormalization scheme.
The GPD in the conformal scheme, $\mathbf{F}_q$, is related to the GPD $F_q$ in the  $\overline{\rm MS}$ scheme by
the U-``rotation''~\footnote{
Note that the kernel $\mathrm{U}(x,x',\xi)$ in Eqs.~\eqref{F-T-C}, \eqref{C-T-C} has to be taken in the
momentum fraction representation. The corresponding expressions can be derived from the results in
Ref.~\cite{Braun:2017cih}, given in position space, but in fact are not needed as
we will find a possibility to avoid this step.}
\begin{align}
\mathbf{F}_q(x,\xi)=[\mathrm{U} F_q](x,\xi)\equiv \int_{-1}^1 \frac{dx'}{\xi} \mathrm{U}(x,x',\xi) F_q(x',\xi)\,,
\label{F-T-C}
\end{align}
and, similarly, for the CF in DVCS~\eqref{C-DVCS}
\begin{align}\label{C-T-C}
C(x/\xi, \mu^2/Q^2 )=\int_{-1}^1 \frac{dx'}{\xi} \mathbf{C}(x'/\xi,\mu^2/Q^2) \mathrm{U}(x',x,\xi)\,.
\end{align}
The ``rotated'' light-ray operator $\mathbf{O}(z_1,z_2)$ in $d=4$ satisfies the RGE
\begin{align}\label{RGOU}
\Big(\mu{\partial_\mu}
+  \mathbf{H}(a_s)
\Big)[\mathbf{O}(z_1,z_2)]=0\,.
\end{align}
Looking for the operator $\mathrm{U}$ in the form
\begin{align}
   \mathrm{U} = e^{\mathbb{X}}\,, \qquad \mathbb{X}(a_s)  =  a_s \mathbb{X}^{(1)}+ a_s^2 \mathbb{X}^{(2)}+\ldots\,,
\label{similarity2}
\end{align}
we require that the ``boldface'' generators do not include conformal anomaly terms,
\begin{subequations}
\label{Sbold}
\begin{align}\label{translationbold}
   \mathbf{S}_- &= {S}_-^{(0)}\,,
\\
\label{dilatationbold}
   \mathbf{S}_0\, &= {S}_0^{(0)} -\epsilon_\ast + \frac12\mathbf{H}\,,
\\
   \mathbf{S}_+   &= {S}_+^{(0)} + (z_1+z_2)\left(-\epsilon_\ast+ \frac12  \mathbf{H}\right)\,,
\label{special-conformalbold}
\end{align}
\end{subequations}
where
\begin{align}
 {S}_-^{(0)} = -\partial_{z_1} -\partial_{z_2}\,, &&
{S}_0^{(0)} = z_1\partial_{z_1} + z_2\partial_{z_2}+2\,, &&
 {S}_+^{(0)} =  z_1^2\partial_{z_1} + z_2^2\partial_{z_2}+2(z_1+z_2)\,,
\label{Scanonical}
\end{align}
are the canonical generators.
Explicit expressions for $\mathbb{X}^{(1)}$ and  $\mathbb{X}^{(2)}$ are given in  \cite{Braun:2017cih}.

With this choice, the generators $\mathbf{S}_\alpha$ on the subspace of the eigenfunctions of the operator $\mathbf{H}$
with a given anomalous dimension $\gamma_N$  take the canonical form with shifted conformal spin
\begin{align}
S_+(\gamma_N)\equiv S_+^{(0)}+(z_1+z_2)\left(-\epsilon_\ast+\frac12\gamma_N\right)\,,
\end{align}
and the eigenfunctions of the rotated kernel,
$\mathbf H\boldsymbol \Psi_{Nk}=\gamma_N \boldsymbol \Psi_{Nk}$, can be
constructed explicitly~\cite{Braun:2011dg}:
\begin{align}
\boldsymbol \Psi_{Nk}=\mathrm{U}\,\Psi_{Nk}\sim
\left(S_+(\gamma_N)\right)^k z_{12}^{N-1}
= z_{12}^{N-1} \frac{\Gamma[2j_N+k]}{\Gamma[j_N]\Gamma[j_N]} \int_0^1\!du\, (t\bar t)^{j_N-1}\, (z_{21}^u)^k.
\end{align}
For the forward matrix element of the light-ray operator one obtains, in our normalization,
\begin{align}
\langle p|[\mathcal O(z_1,z_2)]|p\rangle =\sum_N \frac{i^{N-1}}{(N-1)!}  z_{12}^{N-1}\, \langle p|[\mathcal O_{N}]|p\rangle,
\end{align}
and for the rotated light-ray operator
\begin{align}
\langle p|[\mathbf O(z_1,z_2)]|p\rangle
    =   \sum_N \frac{i^{N-1}}{(N-1)!}  z_{12}^{N-1}\, \sigma_N \,\langle p|[\mathcal O_{N}]|p\rangle\,,
\label{LR-rotated}
\end{align}
where $\sigma_N$ are the eigenvalues of $\mathrm{U}$:
\begin{align}
   \mathrm{U}\, z_{12}^{N-1} = \sigma_N z_{12}^{N-1}\,, \qquad \sigma_N = \sigma_N(a_s)\,.
\label{eq:sigmaN}
\end{align}
The generalization of this expansion to include off-forward matrix elements is completely fixed by conformal algebra
and effectively amounts to the operator relation
\begin{align}
\mathbf{O}(z_1,z_2) = \sum_{Nk} \frac{i^{N-1}}{(N-1)!} \sigma_N \,
a_{Nk} (S_+(\gamma_N))^{k} z_{12}^{N-1} \partial_+^k \mathcal {O}_{N}(0)\,,
\label{LROPE-1}
\end{align}
where
\begin{align}\label{aNk-def}
a_{Nk}=\frac{\Gamma(2j_N)}{k!\Gamma(2j_N+k)}\,, &&  j_N = N + 1 -\epsilon_\ast +\frac12\gamma_N(a_s)\,.
\end{align}
This expression can be derived applying $\partial_+$ to Eq.~\eqref{LR-rotated}.
On the one hand, taking a derivative amounts to a shift $k\to k-1$. On the other hand,
it corresponds to an application of $(- S_-)$ and using the commutation relation
$ S_-S_+^kz_{12}^{N-1} = - k (2j_N+k-1) S_+^{k-1} z_{12}^{N-1}$ one gets a
recurrence relation $a_{N,k-1} = k(2j_N+k-1) a_{N,k}$. The overall normalization
(function of $N$) is fixed by the condition $a_{N,k=0} =1$.

Taking a matrix element of Eq.~\eqref{LROPE-1} between states with fixed momenta and using that $\langle p'|\partial_+^k \mathcal
O(0)|p\rangle = (i \Delta_+)^k \langle p'|\mathcal O(0)|p\rangle $ one obtains
\begin{align}\label{matrix-el-O}
\langle p'|\mathbf{O}(z_1,z_2)|p\rangle &=
\sum_{N}\frac{i^{N-1}}{(N-1)!}  \sigma_N\, \langle p'|\mathcal {O}_N (0)|p\rangle\sum_{k=0}^\infty\!
 a_{Nk} (i\Delta_+)^k S^k_+(\gamma_N) z_{12}^{N-1}\,.
\end{align}
The sum over $k$ can be evaluated with the help of Eq.~(B.10) in Ref.~\cite{Anikin:2013yoa}  (for the special case $n=2)$:
\begin{align}\label{A45}
\sum_{k=0}^\infty (i\Delta_+)^{N-1+k}\,a_{Nk}S_+^k(\gamma_N)z_{12}^{N-1} &=
\frac12 \omega_N\, (-1)^{N-1}
\int_{-1}^1 \! dx\,e^{- i\xi P_+ (z_1+z_2-xz_{12})} \,P^{(\lambda_N)}_{N-1}(x)\,,
\end{align}
where
\begin{align}
P^{(\lambda_N)}_{N-1}(x)=\left(\frac{1-x^2}4\right)^{\lambda_N-\frac12} C_{N-1}^{\lambda_N}(x)\,, &&
 \lambda_N = \frac32 -\epsilon_\ast + \frac12 \gamma_N(a_s)\,,
\label{lambdaN}
\end{align}
$C^{\lambda}_{N}$ are Gegenbauer polynomials,  and
\begin{align}
 \omega_N & =
  \frac{(N-1)!\, \Gamma(2j_N)\Gamma(2\lambda_N)}{\Gamma (\lambda_N+\frac12)\Gamma(j_N)\Gamma(N-1+2\lambda_N)}\,.
\label{omegaN}
\end{align}
Using this representation and the parametrization of the matrix element in Eq.~\eqref{fN(xi)} we obtain
\begin{align}\label{matrix-el-Op}
\langle p'|\mathbf{O}(z_1,z_2)|p\rangle &=
P_+ \sum_{N}\frac{\sigma_N f_N(\xi)}{(N-1!}
\left(\frac{1}{2\xi}\right)^{N-1}
\frac12 \omega_N\,
\int_{-1}^1 \! dx\,e^{- i\xi P_+ (z_1+z_2 - xz_{12})} \,P^{(\lambda_N)}_{N-1}(x)\,.
\end{align}
This expression should be matched to the definition of the GPD in the rotated scheme
\begin{align}
\langle p'|\mathbf O(z_1,z_2)|p\rangle
= 2 P_+ \int_{-1}^1\! dx\, e^{-iP_+ [z_1(\xi-x)+z_2(x+\xi)]} \mathbf F(x,\xi,t)\,.
\label{eq:Frotated}
\end{align}
Changing variables $x\to x/\xi$ one can bring the exponential factor in Eq.~\eqref{matrix-el-Op} to the same form as in
Eq.~\eqref{eq:Frotated} and then try to interchange the order of summation and integration to obtain the answer for the GPD as a series in
contributions of local conformal operators. Attempting this one would find, however, that $\mathbf F(x,\xi,t)$ vanishes outside the
ERBL region $|x|\leq \xi$, which is certainly wrong. This problem is well known and is caused by non-uniform convergence of a sum
representation for GPDs in the DGLAP region $\xi < |x|$. It can be avoided, however, because the CF in DVCS only depends on the
ratio $x/\xi$ so that for our purposes we can set $\xi=1$ and eliminate the DGLAP region completely. In this way we obtain
\begin{align}
\label{Frotated2}
\mathbf{F}(x,\xi =1)=\frac1{4}\sum_{N} \frac{\sigma_N\, \omega_N}{2^{N-1} (N-1)!} \,f_{N}(\xi=1)
P^{(\lambda_N)}_{N-1}(x)
\,,
\end{align}
and the DVCS (vector) amplitude is then given by
\begin{align}
  V(\xi=1,Q^2) &= \int_{-1}^1 dx\,C(x,Q^2) F_q(x,\xi=1) = \int_{-1}^1 dx\,\mathbf C(x,Q^2) \mathbf F_q(x,\xi=1)
\notag\\&=
\frac1{4}\sum_{N} \frac{\sigma_N\, \omega_N}{2^{N-1} (N-1)!} \,f_{N}(\xi=1)
\int_{-1}^1 dx\,\mathbf C(x,Q^2) P^{(\lambda_N)}_{N-1}(x)\,.
\end{align}
On the other hand, from the conformal OPE \eqref{V-COPE}
\begin{align}\label{V-COPE-1}
V(\xi=1,Q^2) &=%
\sum_{N}
f_N(\xi=1)\left(\frac{1}{2}\right)^N C_{1}\left(N,\frac{Q^2}{\mu^2},a_s,\epsilon_\ast\right)\,
\frac{\Gamma(\dhalf-1)\Gamma(2j_N)}{\Gamma(j_N)\Gamma(j_N+\dhalf-1)}\,.
\end{align}
Comparing the coefficients in front of $f_N(\xi=1)$ for these two representations  we obtain
\begin{align}\label{T-OPE-eq}
\int_{-1}^1\!{dx}\, \mathbf{C}(x,Q^2,a_s) P_{N-1}^{(\lambda_N)}(x) =
 C_{1}\left(N,\frac{Q^2}{\mu^2},a_s,\epsilon_\ast\right)
    \frac{2 \Gamma(\dhalf-1)\Gamma(\lambda_N+\frac12) \Gamma(N-1+2\lambda_N)}{\sigma_N\,
\Gamma(2\lambda_N)\Gamma(j_N+\dhalf-1)}\,,
\end{align}
where $d=4-2\epsilon_\ast$ and $\sigma_N$ \eqref{eq:sigmaN} are the eigenvalues of $\mathrm{U}$ on $z_{12}^{N-1}$.
It remains to solve this equation to obtain an explicit expression for the CF in momentum fraction space
and, as the last step, to go over from the ``rotated'' to the $\overline{\rm MS}$ scheme.  In the remaining part of this section
we outline the general procedure for this calculation.

To leading order (tree level) everything is simple. To this accuracy $\gamma_N = 0$, $\lambda_N=3/2$, $j_N = N+1$
and
$C^{(0)}_{1}\left(N\right)=1$ for even $N$ and zero otherwise.
Eq.~\eqref{T-OPE-eq} reduces to
\begin{align}
\int_{-1}^1 dx \,   \mathbf{C}^{(0)}(x) \left(\frac {1-x^2} 4 \right) C_{N-1}^{(3/2)}\left(x\right)=1\,, && N=2,4,\ldots
\,,
\end{align}
and is solved by
\begin{align}
 C^{(0)}(x)= \mathbf{C}^{(0)}(x) = \frac1{1-x}-\frac1{1+x}\,,
\end{align}
in agreement with Eq.~\eqref{CF-oneloop}.
The problem is that beyond the leading order $\lambda_N$ depends on $N$ in a nontrivial way and the functions $P_{N-1}^{(\lambda_N)}\left(x\right)$ are not orthogonal
for different $N$ with some simple weight function. Note, however, that
they are eigenfunctions of the (exact) ``rotated `` evolution kernel
\begin{align}
   \int dx' \, \mathbf{H}(x,x') \, P^{(\lambda_N)}_{N-1}(x') = \gamma_N P^{(\lambda_N)}_{N-1}(x)
   \,,
\end{align}
and also eigenfunctions of the ``rotated'' $SL(2)$  Casimir operator.

This property suggests the following ansatz for the CF:
\begin{align}\label{CK-ansatz}
\mathbf{C}(x)=  \int_{-1}^1 dx'\, C^{(0)}(x') K(x',x)\,,
\end{align}
where $K(x,x')$ is the kernel of a certain $SL(2)$-invariant operator, $[K,\mathbf S_{\pm,0}]=0$ (in momentum
representation). Since the polynomials $P_{N-1}^{(\lambda_N)}\left(x\right)$ are eigenfunctions of the quadratic Casimir operator,
they are also eigenfunctions of {\it any} $SL(2)$-invariant operator, in particular
\begin{align}\label{KPN}
\int dx'\,K(x',x)\,P_{N-1}^{(\lambda_N)}(x') = K(N)\,P_{N-1}^{(\lambda_N)}(x)\,.
\end{align}
Using the above ansatz one obtains
\begin{align}\label{CPN}
\int_{-1}^1 dx \,  \mathbf{C}(x)\, P_{N-1}^{(\lambda_N)}(x)
&=
\int_{-1}^1 dx  \int_{-1}^1 dx'\, C^{(0)}(x') K(x',x)\, P_{N-1}^{(\lambda_N)}(x)
\notag\\&=
K(N)\,\int_{-1}^1 dx \,   C^{(0)}(x) P_N(x)
=2 K(N)B(\lambda_N+\tfrac12,\lambda_N-\tfrac12)\,.
\end{align}
Comparing this expression with Eq.~\eqref{T-OPE-eq} we obtain
\begin{align}\label{spectrum-KN}
K(N)=  C_{1}\left(N,\frac{Q^2}{\mu^2},a_s,\epsilon_\ast \right) \sigma_N^{-1}
    \frac{ \Gamma(\dhalf-1)\Gamma(j_N+\lambda_N-\frac12)}{\Gamma(\lambda_N-\frac12)\Gamma(j_N+\dhalf-1)}\,,
\end{align}
i.e., the spectrum of $K$ is given directly in terms of  moments of the DIS CF and the spectrum
of eigenvalues of the rotation operator $\mathrm{U}$.

An $SL(2)$-invariant operator, i.e., an operator that commutes with the generators $\mathbf S_{\pm,0}$
\eqref{Sbold} of $SL(2,\mathbb{R})$ transformations, is fixed uniquely by its spectrum. Therefore, Eq.~\eqref{spectrum-KN}
unambiguously  defines the operator $K$ and by virtue Eqs.~\eqref{CK-ansatz},~\eqref{C-T-C} also the coefficient function
$\mathbf{C}(x)$. In the next two sections we describe this calculation for the one-loop and the two-loop CFs, respectively.

\section{One-loop example}\label{sec:one-loop}

In this section we take $\mu=Q$ as logarithmic terms $\ln^k(\mu/Q)$ in the CF can easily be restored from the evolution equation. To
one-loop accuracy one needs to expand Eq.~\eqref{spectrum-KN} to order $\mathcal{O}(a_s)$ taking into account that $\epsilon_\ast =
-\beta_0 a_s + \ldots$. Since the tree-level CF does not depend on the space-time dimension, the $\epsilon$-dependence in
$C_{1}\left(N,\frac{Q^2}{\mu^2},a_s,\epsilon_\ast\right)$ starts at order $\mathcal{O}(a_s)$ and can be neglected here. Thus we only
need the one-loop result for $C_1(N)$ in physical $d=4$ dimensions, which can be taken from~\cite{Bardeen:1978yd,Vermaseren:2005qc}:
\begin{align}
 C_{1}(N,a_s) &= 1 + a_s C_1^{(1)}(N) + \ldots
\notag\\
 C_1^{(1)}(N) & = C_F\biggl[
        2\*S_1^2(N)
        -2\*S_2(N)
       -2\*\frac{S_1(N)}{N(N+1)}
       +3\*S_1(N)
       +2\*\frac{1}{N^2}
       +3\*\frac{1}{N}
      - 9
      \biggl]\,,
\end{align}
where $S_{k_1\ldots k_n}(N)$ are harmonic sums~\cite{Vermaseren:1998uu}.
We also need the one-loop flavor-nonsinglet anomalous
dimension
\begin{align}
  \gamma_N^{(1)} &= 4 C_F \Big[2 S_1(N)-\frac{1}{N(N+1)}-\frac32\Big]
  \, ,
\label{eq:gamma-1}
\end{align}
and the (one-loop) eigenvalues $\sigma_N = 1+ a_s\sigma_N^{(1)}+\ldots$ of the rotation matrix
$\mathrm{U} = \II + a_s \mathbb{X}^{(1)}+\ldots $ in Eq.~\eqref{similarity2}.
This is the only new element that requires a calculation.
From Ref.~\cite{Braun:2017cih}
\begin{align}\label{eq:X1}
[\mathbb{X}^{(1)} f](z_1,z_2) & = 2C_F \int d\alpha\,
 \frac{\ln \alpha} \alpha [2f(z_1,z_2)-f(z_{12}^\alpha,z_2)-f(z_1,z_{21}^\alpha)]\, ,
\end{align}
In order to calculate $\sigma_N$ we take $f(z_1,z_2) = z_{12}^{N-1}$ and get
\begin{align}
  \mathbb{X}^{(1)} z_{12}^{N-1} = 4 C_F \int d\alpha \frac{\ln \alpha} \alpha [1-\bar\alpha^{N-1}] z_{12}^{N-1}
~\equiv~ \sigma^{(1)}_N \, z_{12}^{N-1}\, .
\end{align}
where
\begin{align}
\sigma^{(1)}_N = - 2C_F\Big[S_1^2(N-1)+S_2(N-1)\Big]\, .
\label{sigmaN(1)}
\end{align}
Collecting everything we obtain
\begin{align}
K(N) &= 1 + a_s K^{(1)}(N) + \ldots\,,
\notag\\
K^{(1)}(N)  &=  C_1^{(1)}(N) - \sigma^{(1)}_N + \frac12 \gamma^{(1)}_N\,S_1(N+1)
\notag\\&= 8C_F\biggl\{  S_1^2(N) -\frac{S_1(N)}{N(N+1)}+\frac58\frac1{N(N+1)}+\frac14\frac1{N^2(N+1)^2}-\frac98
\biggr\}
\notag\\
&= 2C_F\biggl\{ \left(\bar\gamma^{(1)}_N+\frac32\right)^2+\frac52\frac 1{N(N+1)}-\frac92 \biggr\}\,,
\end{align}
where $\bar\gamma^{(1)}_N$ in the last line is the one-loop anomalous dimension \eqref{eq:gamma-1} stripped of the
color factor, $\gamma_N^{(1)} = 4 C_F \bar\gamma^{(1)}_N$. Note that the asymptotic expansion of the anomalous dimension
$\bar\gamma^{(1)}_N$ and therefore also $K^{(1)}(N)$ at large $j_N$ is symmetric under the substitution $j_N \to 1- j_N$, alias $N\to
-N-1$ (reciprocity relation). We will find that this relation holds to two-loops as well, in agreement with the general
argumentation in~\cite{Basso:2006nk,Alday:2015eya,Alday:2015ewa}, see the next Section.

As already mentioned, a $SL(2)$-invariant operator is completely determined  by its spectrum. It  is easy to do in the
case under consideration, because an invariant operator with eigenvalues $\bar\gamma^{(1)}_N$ is, obviously, the one-loop evolution
kernel, and $1/(N(N+1))$ is nothing else but the inverse Casimir operator. The corresponding explicit expressions in position space
are well known:
\begin{align}
  \widebar{\mathbb H}^{(1)} z_{12}^{N-1}&=
\Big[\widehat {\mathcal H} -\mathcal H_+ -\frac32\Big] z_{12}^{N-1}
 = \bar\gamma_N^{(1)} z_{12}^{N-1}\,,
\notag\\
 \mathcal H_+\, z_{12}^{N-1} &= \frac{1}{N(N+1)} z_{12}^{N-1}\,,
\label{eq:H1}
\end{align}
where
\begin{align}
[\mathcal H_+ f] (z_1,z_2) & =\int_0^1d\alpha \int^{\bar\alpha}_0 d\beta \, f(z_{12}^\alpha,z_{21}^\beta)\,,
\notag\\
[\widehat {\mathcal H} f](z_1,z_2) &=
\int_0^1\frac{d\alpha}{\alpha}\Big[2f(z_1,z_2)-\bar\alpha\big(f(z_{12}^\alpha,z_2)+f(z_1,z_{21}^\alpha)\big)\Big]\,.
\label{eq:position-kernels}
\end{align}
These kernels commute with the canonical generators $S^{(0)}_{\pm,0}$.

Thus the operator $K^{(1)}$ can be written as
\begin{align}\label{K1-kernel}
K^{(1)}=2C_F \biggl[ \left(\widebar {\mathbb H}^{(1)}+\frac32\right)^2 + \frac52 \mathcal H_+  - \frac92\biggr]\,.
\end{align}
The same expression holds in momentum fraction space, apart from that the kernels have to be taken in the corresponding
representation.

For the one-loop example considered here the transformation from position to momentum fraction space
is not difficult to do and the results are available from Ref.~\cite{Braun:2009mi}:
\begin{align}
\mathcal H_+(\var',\var)&=\theta(\var-\var')\frac{\var'}{\var}+\theta(\var'-\var)\frac{1-\var'}{1-\var}\,,
\notag\\
\widehat{\mathcal H}(\var',\var)&=-\theta(\var-\var')\frac{\var'}{\var}\left[\frac1{\var-\var'}\right]_+
+\theta(\var'-\var)\frac{1-\var'}{1-\var}\left[\frac1{\var-\var'}\right]_+
-\delta(\var-\var') \Big(\ln \var+\ln\bar \var\Big)\,,
\end{align}
where $\var$, $\var'$ are rescaled momentum fractions, $\var= (1-x)/2$, and
the plus distribution is defined as
$$
\left[\frac1{\var-\var'}\right]_+ f(\var)=\frac{f(\var)-f(\var')}{\var-\var'}\,.
$$
It remains to calculate the convolution of $K^{(1)}(x,x')$ with the leading-order CF \eqref{CK-ansatz}
and ``rotate'' the result to the $\overline{\rm MS}$ scheme:
\begin{align}
 C^{(1)}(x)= \mathbf{C}^{(1)}(x) + \int_{-1}^1 \!{dx'} \, C^{(0)}(x') \mathbb{X}^{(1)}(x',x)\,.
\end{align}
The one-loop rotation kernel in momentum fraction space $\mathbb{X}^{(1)}(x,x')$ can also be found explicitly,
\begin{align}
[\mathbb{X}^{(1)} f](\var')&\equiv  \int_{0}^1 \!{d\var}\,\mathbb{X}^{(1)}(\var',\var) f(\var)
~=~
2C_F \biggl[\int_{\var'}^1\frac{d\var}{\var} \ln\Big(1-\frac{\var'}{\var}\Big)\frac{ \var' f(\var')
- \var f(\var)}{\var-\var'}
\notag\\&\quad
+\int_0^{\var'}\frac{d\var}{\bar \var}
\ln\Big(1-\frac {\bar \var'}{\bar \var}\Big)\frac{ \bar \var' f(\var')-\bar \var f(\var)}{\bar \var-\bar \var'}
-\frac12 (\ln^2 \var'+\ln^2\bar \var')f(\var')
\biggr]\,.
\label{eq:X1-momentum}
\end{align}
Here, as above, $\var=\frac12(1-x)$, $\var'=\frac12(1-x')$  are rescaled momentum fractions.
Collecting all terms one reproduces after some algebra the well-known expression for the one-loop CF
in Eq.~\eqref{CF-oneloop}.

Beyond one loop, the last part of this strategy --- restoration  of momentum fraction kernels from the
known position space results and taking the remaining convolution integrals --- becomes impractical
because of very complicated expressions. It can be avoided, however, using the following
algorithm.

Let $f(x)$ be a function of the momentum fraction $x$ so that its position space analogue is
\begin{align}
f(z_1,z_2)=\int_{-1}^1 dx\, e^{-iz_1(1-x)-iz_2(1+x)} f(x)\,.
\end{align}
The convolution of $f(x)$ with the leading order CF $C^{(0)}(x)$ can be rewritten
as a position space integral
\begin{align}
\int_{-1}^1 dx\, C^{(0)}(x)\, f(x)=\int_0^\infty dz\, \Big[f(-iz,0)-f(0,-iz)\Big]\,.
\end{align}
Assume that the invariant operator $K$ in position space can be written in the following  form
\begin{align}\label{bold-k}
[Kf](z_1,z_2) &= \int_0^1d\alpha\int_0^{\bar\alpha} d\beta \,\boldsymbol k(\alpha,\beta) f(z_{12}^\alpha,z_{21}^\beta)\,,
\end{align}
where $\boldsymbol k(\alpha,\beta)$ is a certain weight function.
Then
\begin{align}
\int_{-1}^1\! dx\, C^{(0)}(x)\, [Kf](x)&=
\int dx\, f(x) \int d\alpha d\beta\left(\frac{\boldsymbol k(\alpha,\beta)}{\bar\alpha (1-x)+\beta(1+x)}-(x\leftrightarrow -x)\right)\,.
\end{align}
The (momentum fraction space) convolution of the leading order CF and $K$ can therefore be
written directly in terms of the weight function $\boldsymbol k(\alpha,\beta)$
\begin{align}\label{C-K-int}
\int dx' C^{(0)}(x')K(x',x)=
   \int_0^1 d\alpha\int_0^{\bar\alpha} d\beta
   \left(\frac{\boldsymbol k(\alpha,\beta)}{\bar\alpha (1-x)+\beta(1+x)}-(x\leftrightarrow -x)\right)\,.
\end{align}
If $K$ is given by a product of several kernels of the type~\eqref{bold-k}, then the right hand side
of Eq.~\eqref{C-K-int} can be written as a manyfold integral of the same type, e.g., for $K=K_1 K_2$ one gets
\begin{align}
\int da db\int d\alpha d\beta\left(\frac{\boldsymbol k_1(a,b)\boldsymbol k_2(\alpha,\beta)}{
(\bar\alpha\bar a +\alpha b) (1-x)+(\beta\bar a+\bar\beta b)(1+x)}-(x\leftrightarrow -x)\right)\,,
\end{align}
Integrals of this kind can be evaluated with the help of the Maple HyperInt package by E.~Panzer~\cite{Panzer:2014caa} in terms of
harmonic polylogarithms, see e.g.~\cite{Remiddi:1999ew}. In this way a very time consuming transformation of beyond-one-loop
kernels to the momentum fraction representation can be avoided.

For instance, instead of using the momentum fraction expression for $\mathbb X^{(1)}$ in Eq.~\eqref{eq:X1-momentum},
its convolution with $C^{(0)}(x)$ can be calculated using Eq.~\eqref{C-K-int}
directly from the position space representation in Eq.~\eqref{eq:X1}. This leads to simple integrals
\begin{align}
\int dx'\, C^{(0)}(x')\,\mathbb X^{(1)}(x',x) &
=C_F\int_0^1 d\alpha \frac{\ln\alpha}{\alpha}\left(\frac1 { \var}-\frac1{\bar\alpha \var} +\frac1 { \var}
-\frac1{\var+\bar \var\alpha}\right) -(\var\to \bar \var)
\notag\\&=
C_F\frac 1\var\int_0^1d\alpha \left(-\frac{\ln\alpha} {\bar\alpha} +\frac{\bar
\var}{\var}\frac{\ln\alpha}{1+\alpha{\bar \var}/\var}\right) -(\var\to \bar \var)
\notag\\&=
\frac1\var\Big(\Li_2(1)+\Li_2(-\bar \var/\var)\Big) -(\var\to\bar \var)\,,
\end{align}
where $\var=(1-x)/2$. Beyond one loop, this simplification proves to be crucial.

\section{Two-loop coefficient function }\label{sec:two-loop}

The spectrum of the invariant operator $K(N)$ to two-loop accuracy is obtained by expanding Eq.~\eqref{spectrum-KN} to second order
in the coupling. Since $\epsilon_\ast = \mathcal{O}(a_s)$, to this end one needs the two-loop CF in DIS in
$d=4$, and also terms $\mathcal{O}(\epsilon)$ in the one-loop DIS CF as inputs. The corresponding expressions are available from
Ref.~\cite{Zijlstra:1992qd,Vermaseren:2005qc}. In addition we need to calculate the spectrum of eigenvalues $\sigma_N = 1 + a_s\sigma_N^{(1)}+
a_s^2 \sigma_N^{(2)}+\ldots$ of the rotation operator \eqref{similarity2} to the two-loop accuracy. Explicit expressions for the
corresponding kernels
$\mathbb X$ are collected in Appendix B in Ref.~\cite{Braun:2017cih}~\footnote{
In \cite[Eq.~(B.9)]{Braun:2017cih} there is a typo. The second term in the first equation,
 $\sim \mathbb T^{(1)}$, has to be omitted.}.
The necessary integrals can be done analytically in terms of harmonic sums up to fourth order using computer algebra
packages~\cite{Vermaseren:1998uu,Kuipers:2012rf,Ablinger:2010kw,Ablinger:2014rba,Ablinger:2013cf}. The resulting expressions are rather cumbersome
and we do not present them here. The final expressions for the CFs turn out to be considerably shorter thanks to many cancellations.

The next step is to restore the invariant kernel $K$ from its spectrum. This is not as simple as at one loop,
because the invariant kernel has to commute with deformed $SU(3)$ generators \eqref{Sbold}
(including $\mathcal{O}(a_s)$ terms) rather than  the canonical generators \eqref{Scanonical}.
In other words, we are looking now for the integral operator (with the given spectrum)
with eigenfunctions $P^{(\lambda_N)}_{N-1}(x)$ where $\lambda_N = \frac32 +a_s (\beta_0+ \frac12\gamma_N^{(1)})+\ldots$
rather than $\lambda_N =3/2$. All expressions can of course be truncated
at order $a_s^2$ so that for the second-order contributions to the spectrum, $K(N) = \ldots + a_s^2 K(N)^{(2)}$,
it is sufficient to require canonical conformal invariance.
However, we need to modify the first-order kernel \eqref{K1-kernel} $K^{(1)}\mapsto \mathcal K^{(1)}= K^{(1)}+\delta K^{(1)}$
in such a way that $\mathcal K^{(1)}$ has eigenfunctions  $P^{(\lambda_N)}_{N-1}(x)$, i.e. it commutes with
deformed generators $\mathbf S_\alpha$ in Eq.~\eqref{Sbold} (up to terms $\mathcal{O}(a_s^2)$).
This can be achieved by replacing
\begin{align}\label{K1-tilde}
K^{(1)}=2C_F \biggl[ \left(\widebar {\mathbb H}^{(1)}+\frac32\right)^2 + \frac52 \mathcal H_+  - \frac92\biggr]
\quad\Rightarrow\quad
\mathcal K^{(1)}=2C_F\left(\left(\widetilde{\mathbb H}+\frac32\right)^2 +\frac52\widetilde{\mathcal H}_+ - \frac92\right)\,.
\end{align}
where $\widetilde{\mathbb H} = \widebar {\mathbb H}^{(1)} +\mathcal{O}(a_s)$
is the complete two-loop evolution kernel (up to a normalization and some terms discussed below)
and $\widetilde{\mathcal H}_+= \mathcal H_+  + \mathcal{O}(a_s)$ is the inverse to the deformed Casimir operator,
$\widetilde{\mathcal H}_+\sim [\mathbf J(\mathbf J-1)]^{-1}$ (to the required one-loop accuracy).
The spectrum of eigenvalues of $\mathcal K^{(1)}$ will, of course, differ from the spectrum of $K^{(1)}$,
$\mathcal K^{(1)}(N) = K^{(1)}(N) + \mathcal{O}(a_s)$ and this difference will have to be compensated
by the corresponding change in $K^{(2)}\mapsto \mathcal K^{(2)}$, which is, however, straightforward.

The two-loop evolution kernel can be written as~\cite{Braun:2017cih},
\begin{align}
 \mathbb H(a_s) = a_s {\mathbb H}^{(1)} + a_s^2 {\mathbb H}^{(2)} + \ldots\,,
&&
  {\mathbb H}^{(2)}=\mathbb H^{(2,\text{inv})}+ \mathbb T^{(1)}\Big(\beta_0+\frac12\mathbb H^{(1)}\Big)\,,
\end{align}
where $\mathbb H^{(1)}$ is the one-loop kernel \eqref{eq:H1}, \eqref{eq:position-kernels}, $\mathbb T^{(1)}$ is an integral
operator defined in \cite[Eq.~(C.2)]{Braun:2017cih}, and $\mathbb H^{(2,\text{inv})}$ is a certain canonically
$SL(2)$-invariant operator, i.e. $[\mathbb H^{(2,\text{inv})}, S^{(0)}_{\pm,0}]$ =0. It is easy to see that
\begin{align}
 \Big[\mathbb H(a_s), \mathbb H^{(1)} + a_s \mathbb T^{(1)}\Big(\beta_0+\frac12\mathbb H^{(1)}\Big)\Big] = \mathcal{O}(a_s^2)
\,,
\end{align}
so that they have the same eigenfunctions up to $\mathcal{O}(a_s^2)$. In other words,
by throwing away the canonically invariant part $\mathbb H^{(2,\text{inv})}$ of the two-loop evolution kernel
the eigenfunctions remain the same up to terms  $\mathcal{O}(a_s^2)$ that are not relevant to our accuracy.
Thus, we can replace the full evolution kernel in the expression for $\mathcal{K}_1$ in Eq.~\eqref{K1-tilde}
by its (canonically) non-invariant part
\begin{align}
\widetilde{\mathbb H}=\widebar{\mathbb H}^{(1)}+ a_s\widebar{\mathbb T}^{(1)}\left(\beta_0+\frac12\mathbb H^{(1)}\right)\,,
&&
\widetilde{\mathbb H}\, P_{N-1}^{(\lambda_N)} =\widetilde{\mathbb H}(N)\, P_{N-1}^{(\lambda_N)} +O(a_s^2)\,,
\label{eq:H-tilde}
\end{align}
where $\mathbb T^{(1)} \equiv 4 C_F \widebar{\mathbb T}^{(1)}$ and
\begin{align}
 \widetilde{\mathbb H}(N)=\widebar{\mathbb H}^{(1)}(N)+ a_s\widebar{\mathbb T}^{(1)}(N)\left(\beta_0+\frac12\mathbb H^{(1)}(N)\right) =
 \bar \gamma^{(1)}_N+ a_s\left(\beta_0+ 2 C_F\bar\gamma^{(1)}_N\right)\frac{d }{dN}\bar\gamma^{(1)}_N\,.
\label{eq:H-tilde(N)}
\end{align}

In addition, we need to find the inverse of the Casimir operator
\begin{align}
\mathbf J(\mathbf J-1)&=\mathbf S_+\mathbf S_-+\mathbf S_0(\mathbf S_0-1)
\notag\\&
=J_0(J_0-1)+\big(\partial_1 z_{12}+\partial_2 z_{21}+1\big)\Big(-\epsilon_\ast+\frac12\mathbb H\Big)
    +\Big(-\epsilon_\ast+\frac12\mathbb H\Big)^2\,,
\end{align}
where
\begin{align}
 J_0(J_0-1)& =  S_+^{(0)} S_-^{(0)}+ S_0^{(0)} (S_0^{(0)}-1)
=\partial_1 z_{12}(\partial_2 z_{21}+1)=\partial_2 z_{21}(\partial_1 z_{12}+1)\,.
\end{align}
One can show that to the required accuracy
\begin{align}\label{Casimir-expansion}
[\mathbf J(\mathbf J-1)]^{-1}& =\left[1 -a_s \Big(R_1+R_2+\mathcal  H_+\Big)
\Big(\beta_0+\frac12 \mathbb{H}^{(1)}\Big) + \mathcal O(a_s^2)\right]\mathcal H_+\,,
\end{align}
where $\mathcal H_+$ is defined in Eq.~\eqref{eq:position-kernels} and
\begin{align}
R_1f(z_1,z_2) &=(\partial_1 z_{12}+1)^{-1} f(z_1,z_2)=\int_0^1d\alpha\,\bar\alpha f(z_{12}^\alpha,z_2)\,,
\notag\\
R_2f(z_1,z_2) &=(\partial_2 z_{21}+1)^{-1} f(z_1,z_2)=\int_0^1d\alpha\,\bar\alpha f(z_1,z_{21}^\alpha)\,.
\end{align}
The term $\sim \mathcal H_+^2$ is a canonically invariant operator and can be dropped for the same reasons as
$\mathbb H^{(2,\text{inv})}$ in the evolution kernel. Thus, we choose
\begin{align}
\widetilde {\mathcal H}_+= \mathcal H_+ - a_s \Big(R_1+R_2\Big)\left(\beta_0+\frac12 \mathbb{H}^{(1)}\right)\mathcal H_+\,,
&&
\widetilde{\mathcal H}_+\, P_{N-1}^{(\lambda_N)} =\widetilde{\mathcal H}_+(N)\, P_{N-1}^{(\lambda_N)} +O(a_s^2)\,,
\end{align}
where
\begin{align}
\widetilde {\mathcal H}_+(N)=\frac1{N(N+1)}a_s \left[1-\frac{2}{N+1}\Big(\beta_0 + 2 C_F \bar \gamma^{(1)}_N\Big)\right]\,.
\label{eq:H+tilde(N)}
\end{align}
The terms $\mathcal O(a_s)$ in Eqs.~\eqref{eq:H-tilde(N)} and \eqref{eq:H+tilde(N)} modify the spectrum of eigenvalues
of $\mathcal K^{(1)}$ as compared to $K^{(1)}$ and have to be subtracted from $K^{(2)}(N)$. Note, that the two-loop kernel has three
color structures
\begin{align}
\mathcal K^{(2)}(N)&= \beta_0 C_F\mathcal K^{(2)}_\beta(N)+C_F^2 \mathcal K^{(2)}_P(N) +\frac{C_F}{N_c} \mathcal K^{(2)}_A(N)\,.
\end{align}
and only the $\sim C_F^2$ term is affected by this subtraction. We obtain
%
%
{\allowdisplaybreaks
\begin{align}
\mathcal{K}^{(2)}_{\beta}(N) &= 2\zeta_2\bar\gamma_N^{(1)}
+\frac{10}3\left(\bar\gamma^{(1)}_N+\frac32\right)^2
                            - \left(\frac29+7\zeta_2+\frac 8{N(N+1)}+\frac{2}{N^2 (N+1)^2}
                                \right) \left(\bar\gamma^{(1)}_N+\frac32\right)
   \notag\\
   &\quad
              +2 \zeta_3 -\frac{29}6\zeta_2 +\frac{45}8-\frac2{N^2 (N+1)^2}+\frac{31}{6}\frac1{N(N+1)}
\, ,
\notag\\
\mathcal K^{(2)}_{P}(N) & =\frac12 K_1^2(N)+4\zeta_2 (\bar\gamma^{(1)}_N)^2
+ 4\zeta_3 \left(11+\frac{12}{N(N+1)}\right)
-64\zeta_3S_1
+\frac29\pi^4 -\frac{28}3\pi^2 S_1^2
\notag\\
&\quad
+2\pi^2 \left(3+\frac{14}{3N(N+1)}\right)S_1
-2\pi^2\left(\frac49+\frac2{N(N+1)}+\frac1{N^2(N+1)^2}\right)
\notag\\
&\quad
+\frac{6}{N(N+1)}S_{-2}
\!+\!\left(\frac{64}{3}-\frac{14}{N(N+1)}-\frac{8}{N^2(N+1)^2}\right)S_1^2
\notag\\
&\quad
+\!\left(\frac{86}9 -\frac{64}{3N(N+1)}+\frac2{N^2(N+1)^2}+\frac8{N^3(N+1)^3}\right)S_1
\notag\\
&\quad
+\left(\frac{11}8 -\frac{137}{18N(N+1)}-\frac{25}{6N^2(N+1)^2}-\frac2{N^4(N+1)^4}\right)
\, ,
\notag\\[2mm]
\mathcal K^{(2)}_A(N) &=
16S_1\Big(2 S_{1,-2}-S_{-3}\Big)-12 S_{-2}^2-8S_{-4}+16 S_1S_3 + 4\Big(2S_{1,3}- S_4\Big)
-\frac{20 S_3}{N (N+1)}
   \notag\\
&\quad
+\frac{32 \left(S_{-3}-2 S_{1,-2}\right)}{N (N+1)}
+\left(\frac{44}{N^2 (N+1)^2}+\frac{24}{(N-2) (N+3)}+\frac{52}{N (N+1)}+8\right)S_{-2}
   \notag\\
&\quad
+\frac{32}{3} S_1^2
+\left(-\frac{8}{N^3 (N+1)^3}-\frac{8}{N^2(N+1)^2}-\frac{86}{3 N (N+1)}+\frac{52}{9}\!\right) S_1
 \notag\\
&\quad
 +\frac{20}{3 N^2 (N+1)^2}-\frac{59}{9 N (N+1)}+\frac{18}{(N-2) (N+3)}-\frac{35}{4} +\left(\frac{50}{N (N+1)}+54\right) \zeta_3
\notag  \\
&\quad
-\frac{\pi ^4}{9}
-36 \zeta_3 S_1  -\frac{2 \pi ^2 S_1}{N (N+1)}
+\pi ^2 \left(\frac{4}{3 N^2 (N+1)^2}+\frac{2}{3 N (N+1)}-\frac{10}{9}\right)
\, ,
\end{align}}
%
%
where $S_{\vec{m}}\equiv S_{\vec{m}}(N)$. It can be checked that these expressions satisfy the reciprocity
relation~\cite{Basso:2006nk,Alday:2015eya,Alday:2015ewa}: their asymptotic expansion at $N\to\infty$ is symmetric under the
substitution $N\to -N-1$.%
\footnote{The term $\sim S_{-2}$ in $\mathcal K^{(2)}_{P}$ may seem curious
as such harmonic sums do not appear in two-loop planar diagrams.
This term  arises from the two-loop
contribution to $\sigma_N$ \eqref{eq:sigmaN} and is related to the particular choice of the
``rotation'' operator $\mathrm U$ in \cite{Braun:2017cih}. The definition of $\mathrm U$  involves
certain ambiguity which does not affect, of course, the final answer.}

It remains to find an invariant operator $\mathcal K^{(2)}$ with the spectrum $\mathcal K^{(2)}(N)$.  This is not very hard to do
since we only need canonical $SL(2)$-invariance, i.e. the operators in question have to be diagonal in the basis of
$P_{N-1}^{(3/2)}(x)$. The $SL(2)$-invariant kernels with eigenvalues given by the required harmonic sums are collected in
Appendix~\ref{App:KernelList}. Note, that in difference to anomalous dimensions which grow logarithmically at large N to all orders
in perturbation theory, $\gamma^{(k)}_N \sim \ln N$, the kernel $K(N)$ contains higher powers on the logarithm,  up to $\ln^{2\ell}
N \sim S_1^{2\ell}(N)$, where $\ell$ is the number of loops.

With the invariant operators at hand, the DVCS coefficient function in the rotated scheme is obtained by the convolution
with the leading-order CF
\begin{align}
\mathbf C(x) & =  \int_{-1}^1 dx'\, C^{(0)}(x') \, K(x',x) =
   \int_{-1}^1 dx'\, C^{(0)}(x') \,\Big[\delta(x-x')+ a_s K^{(1)}(x',x) + a^2_s K^{(2)}(x',x)\Big]
\,,
\end{align}
and the CF in the $\overline{\text{MS}}$ scheme (so far still in conformal QCD at the critical point) recovered as
\begin{align}
  C(x) &= \int_{-1}^1 dx'\, \mathbf C (x') \,
 \Big[\delta(x-x') +a_s \mathbb X_1(x',x)+
a_s^2\Big( \frac12 \mathbb X_1^2 +\mathbb X_2\Big)(x',x)+\ldots\Big)\Big]\,.
\end{align}
In both cases one can follow the procedure described in the previous section and avoid a Fourier transformation of the kernels to the
momentum fraction space. All necessary integrals can be computed using the HyperInt package~\cite{Panzer:2014caa} in terms of
the harmonic polylogarithms~\cite{Remiddi:1999ew}. The two-loop CF contains contributions of three color structures
\begin{align}
C^{(2)}_{*}(x)\equiv C^{(2)}(x,a_s,\epsilon_\ast)~=~\beta_0 C_F\, C^{(2\beta)}_{\ast}(x)+C_F^2 C^{(2P)}_{\ast}(x)
+\frac{C_F}{N_c} C^{(2A)}_{\ast}(x)\,.
\end{align}
We obtain
%
%
{\allowdisplaybreaks
\begin{align}\label{C2star}
 C^{(2\beta)}_\ast(x)&=
\frac1 \omega\Big( \mathrm H_{1,0,0}-\frac12\mathrm H_{1,1,0}\Big)
    +\left(\frac{5}{3\omega}-\frac1{\bar \omega}\right) \mathrm H_{0,0}
    -\left(\frac1{\bar \omega}+\frac{7}{3\omega}\right) \mathrm H_{1,0}
    \notag\\
    &\quad -\frac{\zeta_2}{2\omega}\Big(\mathrm H_{1}+\mathrm H_0\Big)+\left(\frac1{12\bar \omega}-\frac{14}{9 \omega}\right)\mathrm H_0
-\frac1{2\omega}\left(\frac{25}{24}+\frac{25}6 \zeta_2 + \zeta_3\right) - (\omega\to \bar \omega)
\, ,
\notag   \\
%
C^{(2P)}_\ast(x) &=
\frac1\omega\Big(6\mathrm H_{0,0,0,0}-\mathrm H_{1, 0, 0, 0}-2\mathrm H_{2, 0, 0}
-\mathrm H_{1, 1, 0, 0}-\mathrm H_{1, 2, 0}-\mathrm H_{2, 1, 0}+\mathrm H_{1, 1, 1, 0}\Big)
\notag\\
&\quad
-\frac1{\bar \omega}\,\mathrm H_{0, 0, 0}
-\left(\frac4\omega -\frac2{\bar \omega}\right)\mathrm H_{1, 0, 0}
+\frac1{\bar \omega}\,\mathrm H_{2, 0}
+\frac2\omega\,\mathrm H_{1, 1, 0}
\notag\\
&\quad
-\left(\frac{13}{2\bar \omega}+\frac{19}{3\omega}\right)\mathrm H_{0, 0}
+\left(\frac3{\bar \omega}+\frac{11}{3\omega}\right)\mathrm H_{1, 0}
+\frac{1}\omega \zeta_2\Big( \mathrm H_{1, 1}-\mathrm H_{2}-\mathrm H_{1, 0}-4\mathrm H_{0, 0}\Big)
\notag\\
&\quad
+\left(
\frac1{\bar \omega}\left(\frac{223}{12}+5\zeta_2-2\zeta_3\right)
+\frac1{\omega}\left(3\zeta_2+16\zeta_3-\frac{32}9\right)
\right)\mathrm H_{0}
\notag\\
&
\quad+\frac1{48 \omega}\Big(701+128\zeta_2+936\zeta_3+72\zeta_2^2\Big) -(\omega\leftrightarrow \bar \omega)
\, ,
\notag   \\
%
C^{(2A)}_\ast(x) &=
6(1-2\omega)\Biggl\{
\mathrm H_{2, 0}-\mathrm H_{3}+\mathrm H_{1, 1, 0}-\mathrm H_{1, 2}
+\zeta_2 \Big (\mathrm H_0+\mathrm H_1\Big)-3\zeta_3\Biggr\}
\notag\\
&\quad
+12\Big(\mathrm H_{1, 0} - \mathrm H_{2} - \mathrm H_{0} - \mathrm H_{1}+\zeta_2\Big)
+\frac 3{\bar \omega}\mathrm H_{0}
+\frac3\omega\mathrm H_{1}
\notag\\
&\quad +
\Biggl\{\frac1\omega\left(12\zeta_3 - \frac32\zeta_2^2 - 
3\zeta_2 -\frac{73}{24}\right)
-\frac3\omega\mathrm H_{2, 0, 0}
      -\left(\frac2 \omega -\frac1{\bar \omega}\right)\mathrm H_{3, 0}
           +\left(\frac4\omega-\frac1{\bar \omega}\right)\mathrm H_{4}
\notag\\
&\quad
            -\left(\frac2\omega-\frac1{\bar \omega}\right)\mathrm H_{2, 1, 0}
          +\left(\frac{3}\omega-\frac2{\bar \omega}\right)\mathrm H_{2,2}
           -\left(\frac2{ \omega}-\frac1{\bar \omega}\right)\mathrm H_{3, 1}
           -\frac5{\bar \omega}\mathrm H_{3}
           +\frac{5}{\bar \omega}\mathrm H_{2, 0}
\notag\\
&\quad
            +\left(\frac1{\bar \omega}\left(\zeta_2-
            5\right)+\frac1\omega \left(\frac{4}3-2\zeta_2\right)\right)\mathrm H_{0, 0}
           -\left(\frac2\omega\left(\zeta_2-1\right)-\frac1{\bar \omega}\left(\zeta_2+
           \frac23\right)
           \right)\mathrm H_{2}
\notag\\
&\quad
+\left(\frac1{\bar \omega}\left(\frac{19}6+ 5\zeta_2-3\zeta_3\right)
+\frac1\omega\left(7\zeta_3 -\frac{16}9\right)
         \right)\mathrm H_{0}
           -(\omega\leftrightarrow \bar\omega)
           \Biggr\}
\, ,
\end{align}}
%
%
where $\var=(1-x)/2$, $\bar \var=(1+x)/2$, and $\mathrm H_{\vec{m}}\equiv \mathrm H_{\vec{m}}(\var)$ are harmonic polylogarithms.
We also note that first two lines in the expression for $C^{(2A)}_\ast(x)$ are entirely due to terms $\sim 1/(N-2)$ in the
corresponding kernel.

Finally, the QCD result in $d=4$ is recovered by adding the $\mathcal{O}(\epsilon)$ correction to the one-loop CF, cf.
Eq.~\eqref{eq:d=4-restored}, which is easy to find by a direct calculation:
\begin{align}
 C^{(2)}(x) & = C_\ast^{(2)}(x) +\beta_0 C^{(1,1)}(x)\,,
\notag\\
C^{(1,1)}(x) &=
   - C_F \frac{1}{2 \omega} \biggl\{
18 - \frac{\pi^2}{4} - \left(5 -\frac{4}{\bar\var} + \frac{\pi^2}{6}\right) \ln \var
- \frac{3}{2} \frac{\var}{\bar\var} \ln^2\var + \frac13 \ln^3\var\biggr\}
-(\omega\leftrightarrow \bar\omega)
\, .
\label{eq:d=4-result}
\end{align}
%
%
Note, that the contribution $\sim\beta_0$ arises in our calculation as a sum of several terms: in the DIS CF, in the rotation matrix
$\mathrm{U}$, and in $C^{(1,1)}(x)$. This contribution can also be calculated directly from the fermion bubble insertion $\sim n_f$ in the one-loop diagrams. We did this calculation and checked that the results agree.
Similar calculations of the $\sim n_f\alpha_s^2$ terms exist in the literature, e.g., for the axial-vector case~\cite{Melic:2001wb}.

The leading double-logarithmic asymptotic  of the CF at $\omega\to 0$ can easily be obtained to two-loop accuracy from the explicit
expressions given above. The result reads
\begin{align}
C(x,a_s)& \simeq\frac1{2\omega}\Big(1+C_F a_s \ln^2\omega+\frac12(C_F a_s)^2\ln^4\omega+\mathcal O(a_s^3)\Big)\,,
\label{eq:threshold}
\end{align}
suggesting that the series exponentiates. This expression does not agree with the resummed formula obtained in
Ref.~\cite{Altinoluk:2012nt}.

\subsection{Restoring the scale dependence}
\label{sec:scale}
The scale-dependent $\sim \ln Q/\mu$ terms in the CF are completely fixed by the RGE.
Since the evolution kernel in the $\overline{\rm MS}$ scheme does not depend on $\epsilon$,
$\mathbb{H}(a_s,\epsilon) = \mathbb{H}(a_s)$, in a generic $d$-dimensional theory
\begin{align}
\Big(\mu\partial_\mu + \beta(a_s,\epsilon)\partial_{a_s}\Big)C\left(Q^2/\mu^2,a_s,\epsilon\right) =
C\left(Q^2/\mu^2,a_s,\epsilon\right)\otimes  \mathbb H(a_s)\,,
\end{align}
where
\begin{align}
    C\otimes\mathbb H = \int_{-1}^1 dx'\, C(x')\, \mathbb H(x',x)\,.
\end{align}
Solving this equation one obtains
\begin{align}
C(\sigma,a_s,\epsilon) & =\Big( C^{(0)}+ a_s C^{(1)}(\epsilon)+a_s^2 C^{(2)}(\epsilon)+\ldots \Big)\otimes
\Big(1 -\frac12 \ln \sigma\, \mathbb H(a_s)+ \frac18 \ln^2\!\sigma\, \mathbb H^2(a_s) +\ldots   \Big)\,
\notag\\
&\quad
-\beta(a_s,\epsilon)\left( -\frac12 C_1(\epsilon)\,\ln \sigma + \frac1{8a_s} \ln^2s\,C_0\otimes \mathbb H(a_s)\right)
+ O(a_s^3,a_s^2\epsilon, a_s\epsilon^2)\,.
\end{align}
Here $\sigma = Q^2/\mu^2$ and $C^{(0)}, C^{(1)}(\epsilon), C^{(2)}(\epsilon)$ are the CFs in $d$ dimensions \eqref{eq:generic-d}
at $\mu^2=Q^2$,
alias $\sigma=1$. Note that the contribution in the second line vanishes at the critical point, $\beta(a_s,\epsilon_\ast)=0$.
For the physical case $d=4$ one obtains
\begin{align}
C(\sigma,a_s,\epsilon=0) & = C^{(0)}+ a_s \biggl(C^{(1)}_\ast -\frac12 \ln \sigma \, C^{(0)}\otimes \mathbb H^{(1)}\biggr)
+a_s^2\biggl\{C_2^{(*)}+\beta_0 C^{(1,1)}
\notag\\&\quad
-\frac12 \ln \sigma \biggl[ C^{(0)}\!\otimes\mathbb H^{(2)}
+2 C^{(1)}\!\otimes \Big(\beta_0+\frac12\mathbb H^{(1)}\Big) \biggr]
+\frac14 \ln^2\!\sigma\, C^{(0)}\!\otimes \mathbb H^{(1)}\Big(\beta_0+\frac12\mathbb H^{(1)} \Big)\!\biggr\}\,.
\end{align}
where the CFs in $d=4$ are related to the ones at critical coupling as
$C^{(1)}(\epsilon = 0)= C^{(1)}_\ast$ and $C^{(2)}(\epsilon = 0)= C_\ast^{(2)}+\beta_0 C^{(1,1)}$, see Eq.~\eqref{eq:d=4-restored}.
In terms of the $\omega$-variable, $\omega = (1-x)/2$,
\begin{align}
C^{(0)} &=\frac 1{2\omega}-\frac1{2\bar\omega}\,,
\notag\\
C_\ast^{(1)} & = C_F\left(\frac1{2\omega} \ln^2\omega -\frac{3}{2\bar\omega} \ln\omega-\frac9{2\omega}\right)
 -\big(\omega\mapsto \bar \omega \big)\,.
\end{align}
The functions $C_\ast^{(2)}$ and $C^{(1,1)}$ are given in Eqs.~\eqref{C2star} and \eqref{eq:d=4-result}, respectively.

Using explicit expressions for the one- and two-loop evolution kernels~\cite{Braun:2017cih} we obtain
\begin{align}
C(x,\sigma ) & =C^{(0)}(x)+ a_s \Big(C_\ast^{(1)}(x) -\frac12 \ln \sigma \, A_1\Big)+a_s^2 \Big(C_\ast^{(2)}(x)
+\beta_0 C^{(1,1)}(x)-\frac12\ln \sigma \, A_2
+\frac14 \ln^2 \sigma \,B_2\Big)\,,
\end{align}
where
{\allowdisplaybreaks
\begin{align}
A_1 &=
-2C_F\left( \frac1\omega\left(\mathrm H_0+\frac{3}{2}\right) -
  \big(\omega\mapsto \bar \omega \big)\right)
\, ,
\notag\\
B_2 &=
-2\beta_0 C_F \frac1\omega\left(\mathrm H_0+\frac{3}{2}\right)
+
4C_F^2\left(\frac1\omega \mathrm H_{1,0}+\frac2\omega \mathrm  H_{0,0}+\left(\frac1{\bar \omega}+\frac3 \omega\right) \mathrm H_0
 +\frac9{ 4\omega}\right)
 -  \big(\omega\mapsto \bar \omega \big)
\, ,
\notag\\
A_2 &=
2\beta_0\, C^{(1)}_\ast -
2\beta_0 C_F \biggl\{
\frac 1 \omega \mathrm H_{1,0}
    + \left(\frac{5}{3\omega}-\frac1{\bar \omega}\right)\mathrm H_{0} + \frac1{\omega}\left(\zeta_2+\frac14\right)
 -  \big(\omega\mapsto \bar \omega \big)
\biggr\}
\notag\\
&\quad
 + \frac{4C_F}{N_c} \biggl\{ \frac1\omega\Big(\mathrm H_{2,0}-\mathrm H_{3}\Big)
+ \left(\frac1\omega\left(
\zeta_2-\frac 2 3\right)+\frac1{\bar \omega}
\right) \mathrm H_{0}
+\frac1{\omega}\left(3\zeta_3-\frac{1}{4}\right)  -  \big(\omega\mapsto \bar \omega \big)
\biggr\}
\notag\\
&\quad
+2C_F^2\biggl\{\frac2 \omega \mathrm H_{1,1,0}
-\frac 6 \omega \mathrm H_{0,0,0}  +\left(\frac1{\bar\omega} -\frac 3 \omega \right)\mathrm H_{0,0}
+\frac4 \omega \mathrm H_{1,0}
\notag\\
&\quad
+\left(\frac1\omega \left(4\zeta_2+\frac{19}3\right) +
\frac1{\bar\omega}\left(2\zeta_2+\frac{19}2\right)\right)\mathrm H_{0}
+ \frac1 \omega \left(3\zeta_2 -2\zeta_3 + \frac{47}4\right)
 -  \big(\omega\mapsto \bar \omega \big)
\biggr\}
\, .
\end{align}}
These expressions represent our final result.

\section{Numerical estimates}\label{sec:numerics}

The numerical results in this section are obtained assuming the photon virtuality $Q^2=4$~GeV$^2$ and the corresponding value
of the strong coupling $a_s(4\,\text{GeV}^2) = \alpha_s(4\,\text{GeV}^2)/(4\pi) = 0.02395$.
The DVCS CF in the ERBL region ($|x|<\xi$) is given directly by the above expressions with an obvious substitution
$x\to x/\xi$ and is obtained by the analytic continuation $\xi \to \xi-i\epsilon$ in the DGLAP region $|x|>\xi$.
We have used the Mathematica package HPL-2.0 by D.~Maitre~\cite{Maitre:2005uu,Maitre:2007kp} for a
numerical evaluation of the  harmonic polylogarithms at complex arguments.

\begin{figure}[t]
\begin{center}
\includegraphics[width=0.47\linewidth]{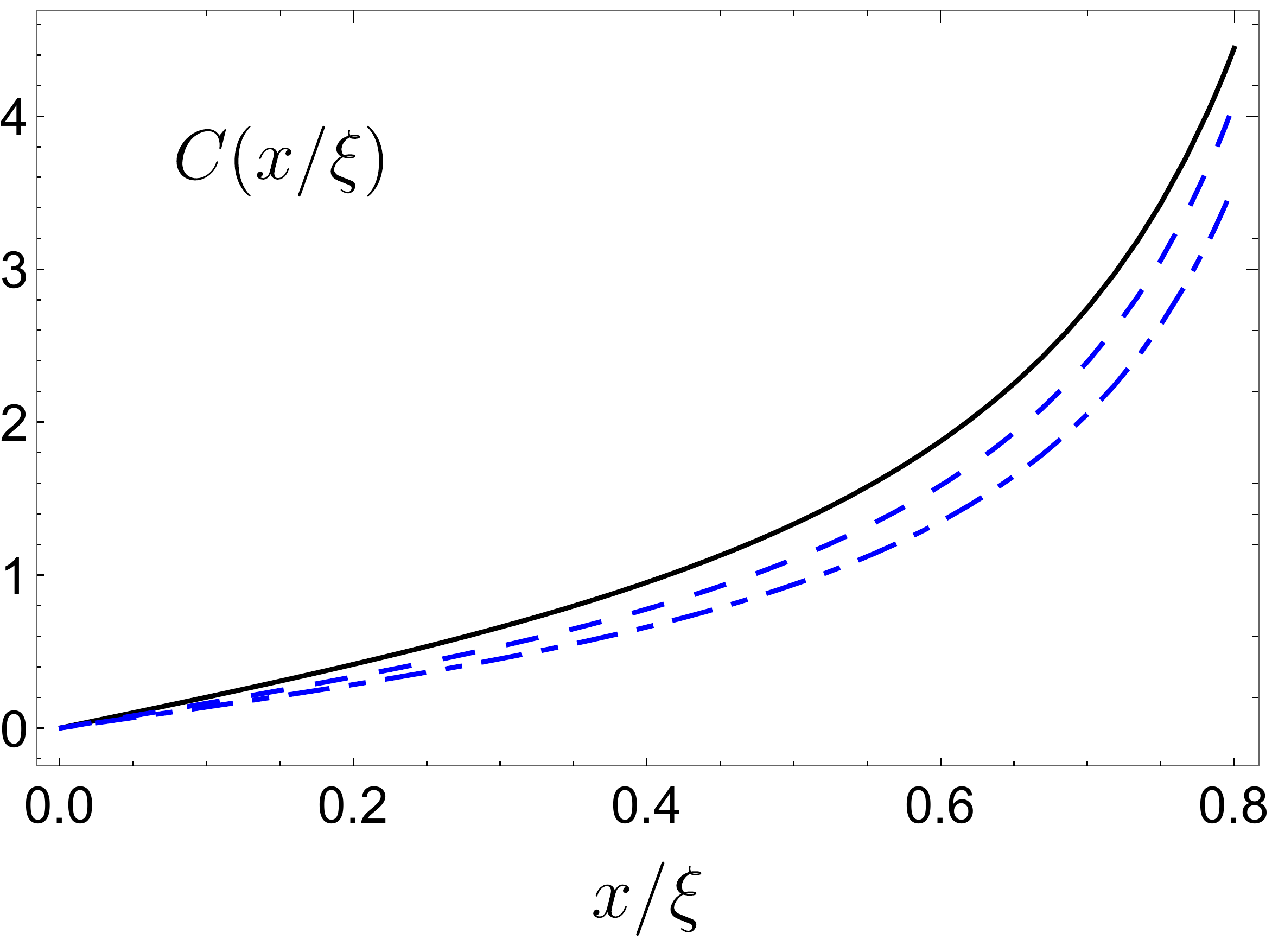}\hskip 5mm
\includegraphics[width=0.49\linewidth]{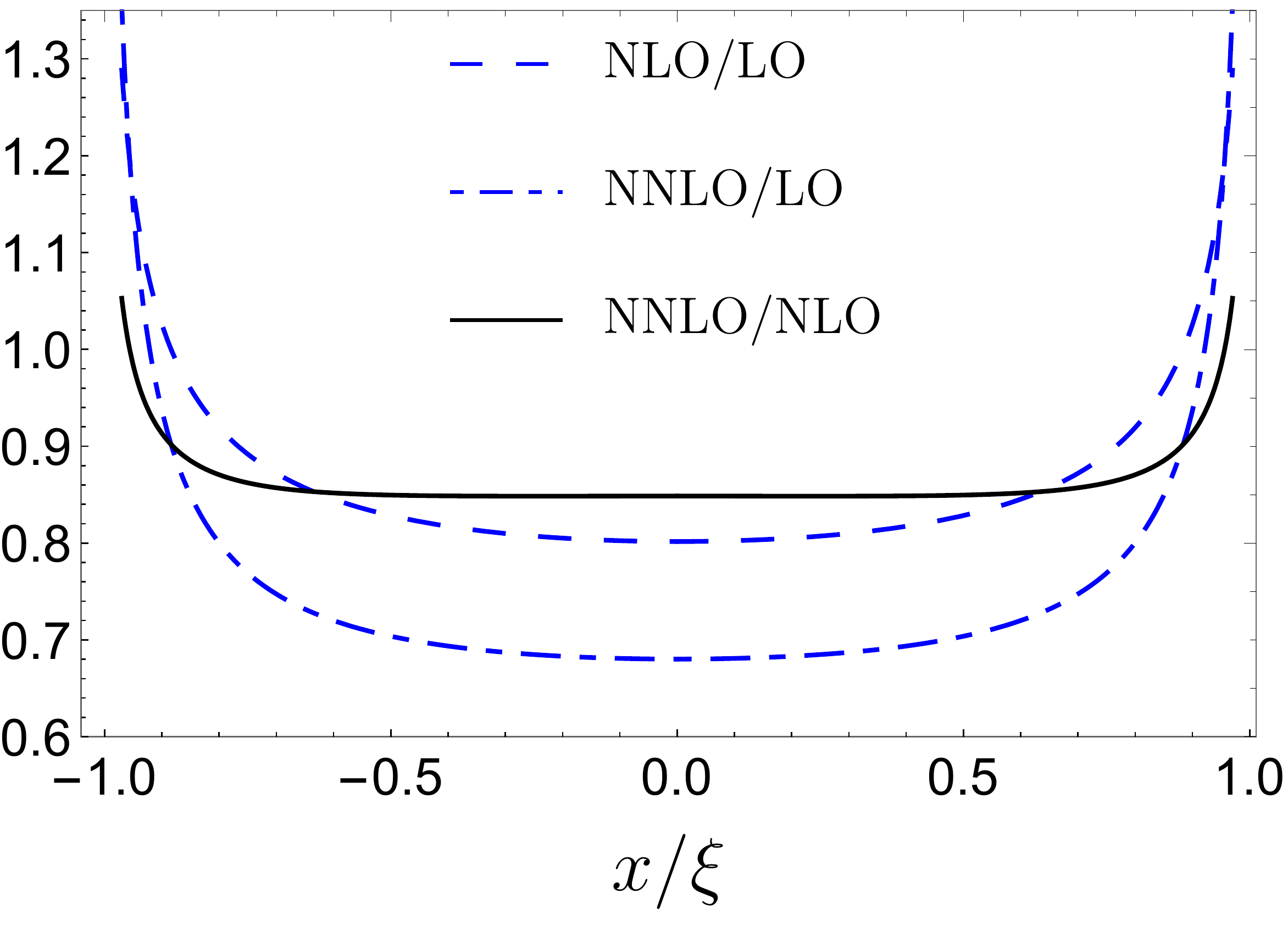}
\end{center}
\caption{The DVCS CF $C(x/\xi)$ in Eqs.~\eqref{C-DVCS}, \eqref{eq:series} at $\mu = Q = 2~\text{GeV}$
in the ERBL region $x<\xi$.
The LO (tree-level), NLO (one-loop) and NNLO (two-loop) CFs are shown by the black solid, blue dashed and
blue dash-dotted curves on the left panel, respectively. The right panel shows the ratios NLO/LO (dashed),
NNLO/LO (dash-dotted) and NNLO/NLO (solid).
}
\label{fig:C-ERBL}
\end{figure}

\begin{figure}[ht]
\begin{center}
\includegraphics[width=0.47\linewidth]{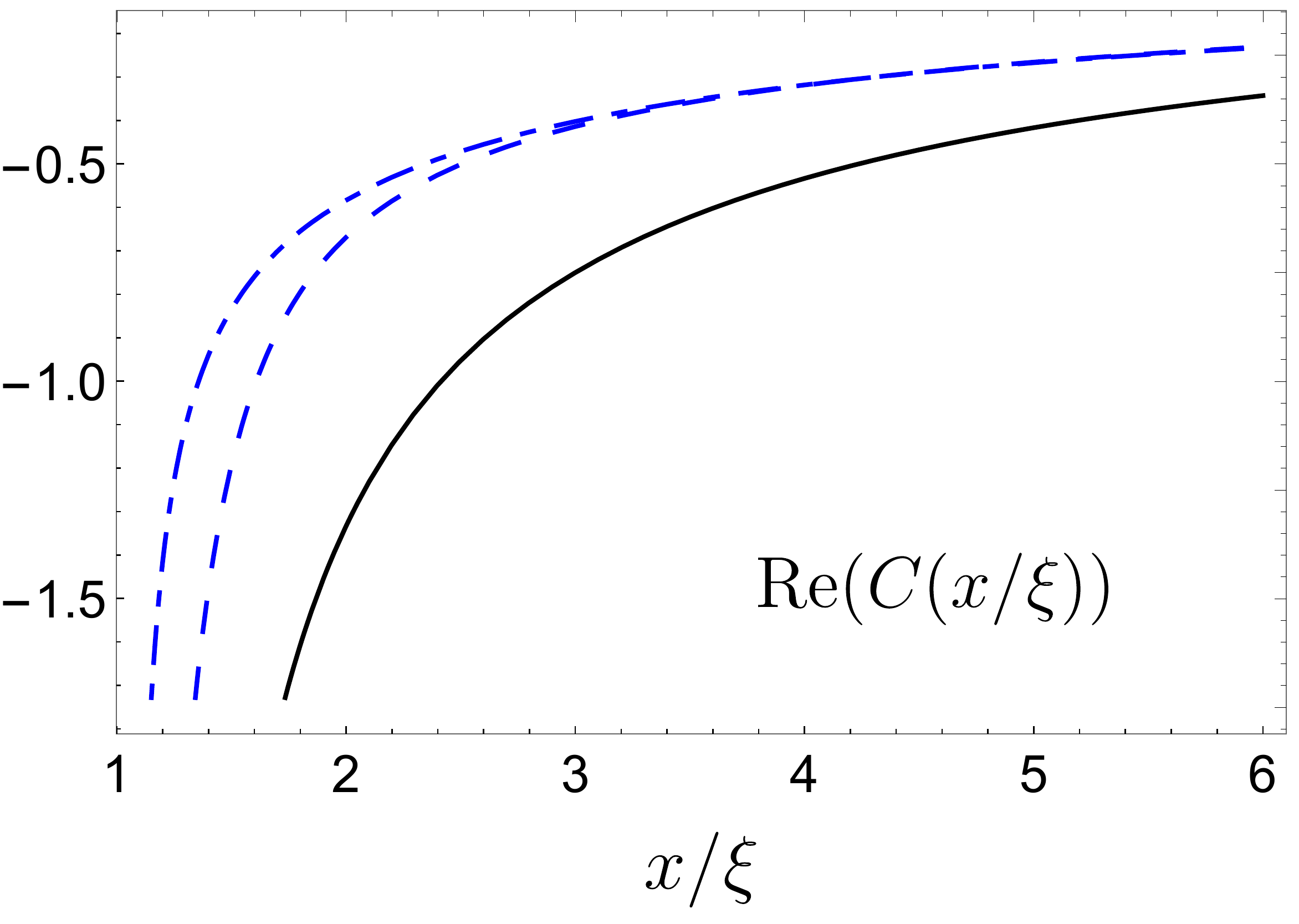}\hskip 5mm
\includegraphics[width=0.48\linewidth]{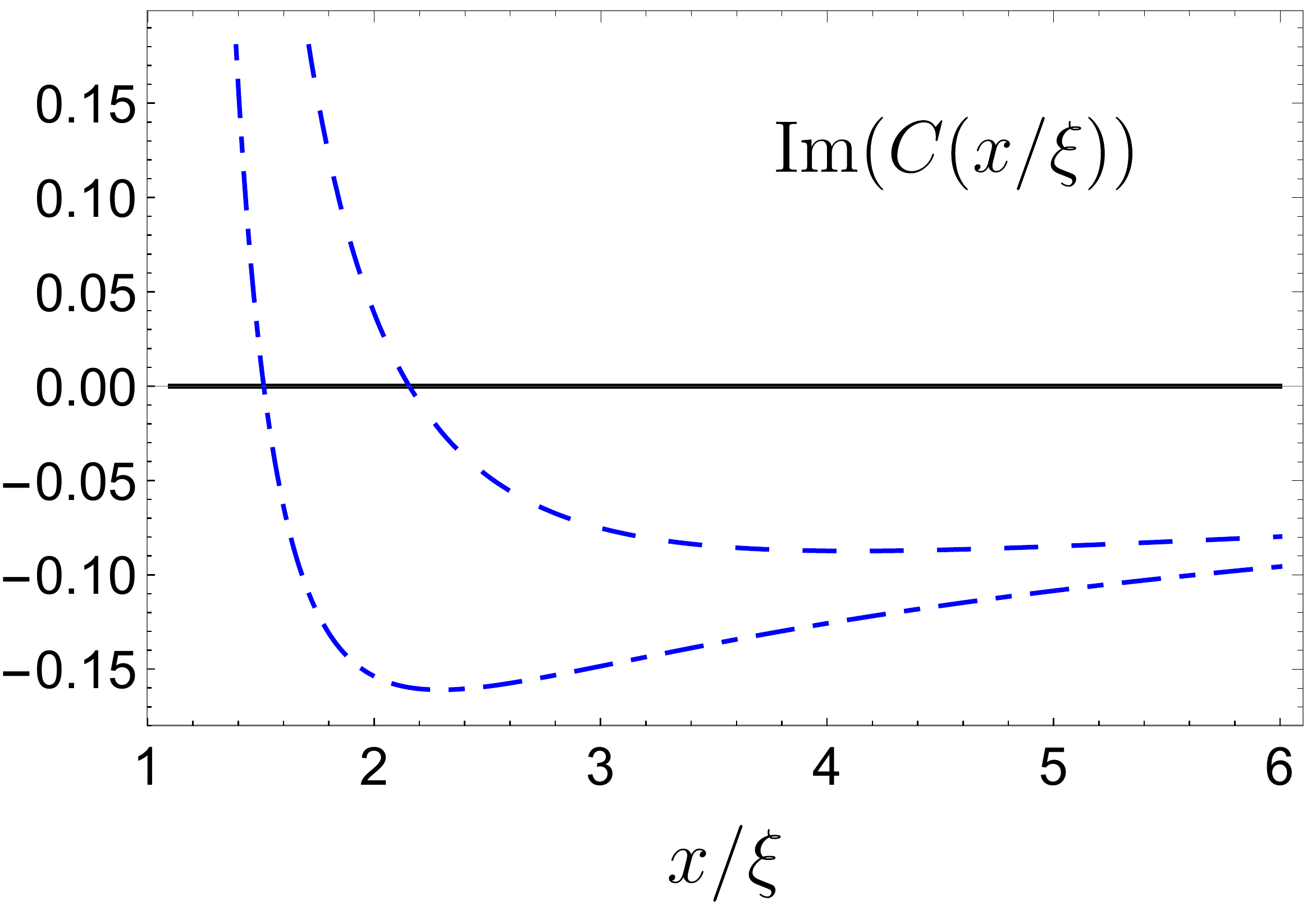}
\end{center}
\caption{The DVCS CF $C(x/\xi)$ in Eqs.~\eqref{C-DVCS}, \eqref{eq:series} at $\mu = Q = 2~\text{GeV}$
analytically continued into the DGLAP region $x>\xi$: real part on the left and imaginary part on the right panel.
The LO (tree-level), NLO (one-loop) and NNLO (two-loop) CFs are shown by the black solid, blue dashed and
blue dash-dotted curves. Note, that imaginary part of the LO CF contains a local term $\sim \delta(x-\xi)$ (not shown).
}
\label{fig:C-DGLAP}
\end{figure}

The results are shown in Figs.~\ref{fig:C-ERBL} and  \ref{fig:C-DGLAP}, respectively. In the first figure,
we also show on the right panel the ratios of NLO and NNLO to the leading order (LO) contribution,
NLO/LO and NNLO/LO, and the ratio NNLO/NLO.
It is seen that the NNLO (two-loop) and NLO (one-loop) contributions to the CF
have the same sign and are negative with respect to the
LO (tree-level) result in the bulk of the kinematic region apart from the end points $|x|\to |\xi|$ where the
loop corrections are positive and dominated by the contributions of threshold double-logarithms~\eqref{eq:threshold}.
We observe that the NNLO contribution is significant. In the ERBL region, it is generally about 10\% of the LO result
(factor two below NLO). In the DGLAP region it is less important and in fact negligible for the real part
at  $x/\xi>2$, and for the imaginary part at $x>4\xi$.


\begin{figure}[ht]
\begin{center}
\includegraphics[width=0.6\linewidth]{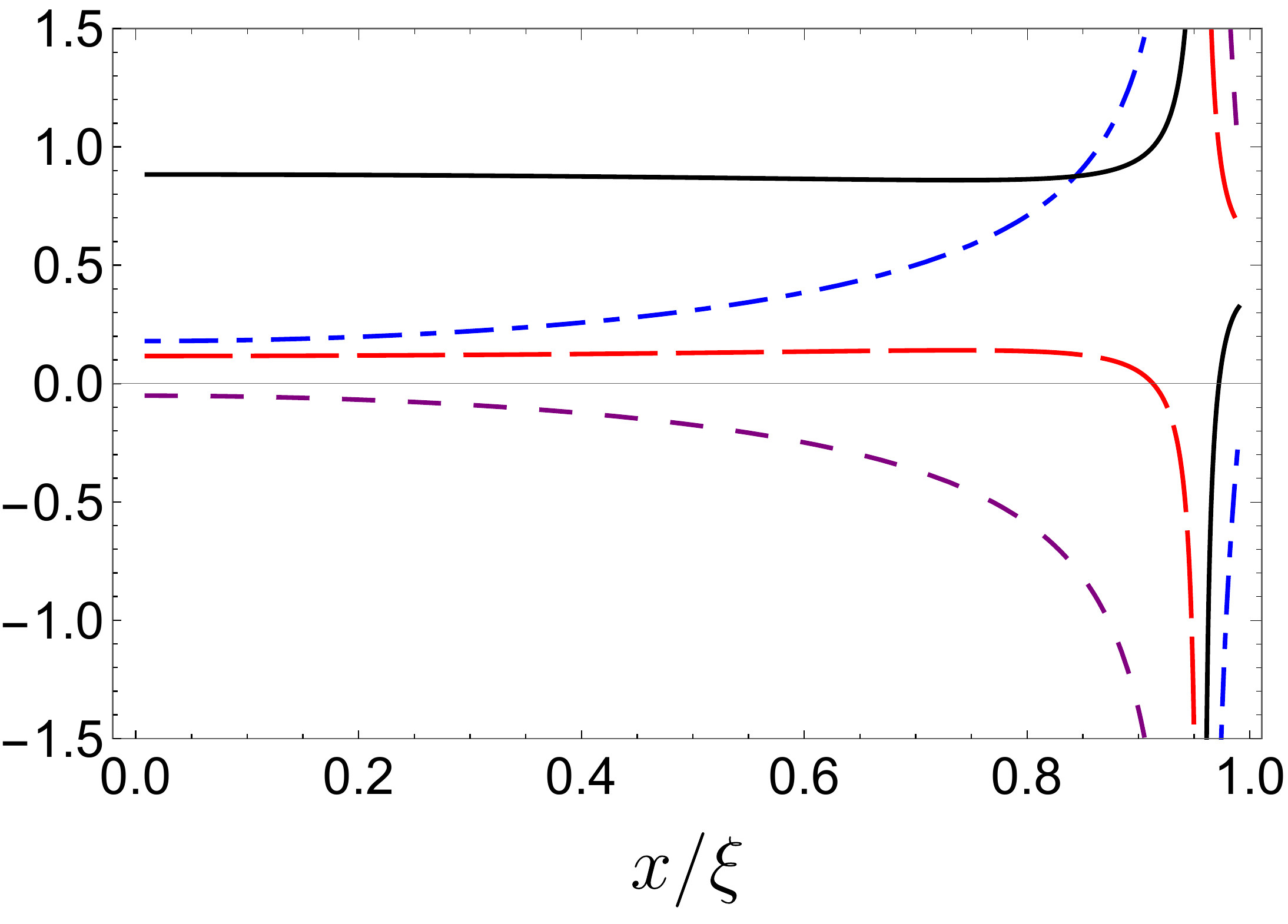}
\end{center}
\caption{Relative contributions of different structures to the NNLO CF: the black line $\beta_0 C^{(1,1)}/C^{(2)}$,
the red line $C_*^{(2)}/C^{(2)}$, the dashed line $\beta_0 C_F C^{(2\beta)}_*/C^{(2)}$ and
the dot-dashed line $ C_F^2 C^{(2P)}_*/C^{(2)}$.
}
\label{fig:C2-decompose}
\end{figure}

The relative size of the contributions of different origin to the NNLO CF  is illustrated in Fig.~\ref{fig:C2-decompose}.
We repeat the definitions for convenience:
\begin{align}
C^{(2)}(x)& = \beta_0 \,C^{(1,1)}(x) + C_\ast^{(2)}(x)\,,
\notag\\
C_\ast^{(2)}(x) &= \beta_0 C_F\, C^{(2\beta)}_{\ast}(x)+C_F^2 C^{(2P)}_{\ast}(x) +\frac{C_F}{N_c} C^{(2A)}_{\ast}(x)\,.
\end{align}
Here $C_\ast^{(2)}(x)$ defines the CF in conformal QCD at the critical point and the  $\beta_0 \,C^{(1,1)}(x)$ term
describes the shift to integer $d=4$ dimension. We show in Fig.~\ref{fig:C2-decompose} the ratio
$\beta_0 C^{(1,1)}/C^{(2)}$ by the black solid, $\beta_0 C_F C^{(2\beta)}_*/C^{(2)}$ by the dashed, and
$C_F^2 C^{(2P)}_*/C^{(2)}$ by the dash-dotted curves, respectively. The  $C_*^{(2)}/C^{(2)}$ ratio is shown by
the red dashed curve; it includes also the contribution of non-planar diagrams $\sim C^{(2A)}_{\ast}$ which is
very small numerically. We observe that $C^{(2)}(x)$ is dominated in almost the entire range of $x$
by the simplest contribution, $\beta_0 \,C^{(1,1)}(x)$, that comes from the $\mathcal{O}(\epsilon)$
correction to the one-loop diagrams. The CF in the conformal theory at $\epsilon = \epsilon_\ast$ is small as the
result of a strong cancellation between the contributions $\sim C_F^2$ of planar diagrams and the term
proportional to the QCD $\beta$ function, $\beta_0 C_F\, C^{(2\beta)}_{\ast}(x)$,  that arises at the critical point
due to the relation $\epsilon_\ast = -\beta_0 a_s +\ldots$.

The dominance of the $\beta_0 \,C^{(1,1)}(x)$ term does not hold for very large $x/\xi \gtrsim 0.95$.
In this region $C^{(2)}$  changes sign so that the representation in Fig.~\ref{fig:C2-decompose} is not very informative.
Asymptotically, for $x/\xi \to 1$, the NNLO CF $C^{(2)}$ is dominated by the Sudakov-type double-logarithmic term
$\sim a_s^2C_F^2\ln^4(1-x/\xi)$ in Eq.~\eqref{eq:threshold} that is part of the $C^{(2P)}_{\ast}(x)$ contribution.
This only happens very close to the end-points, however, for $x/\xi \gtrsim 0.99$.

Physical observables in DVCS are Compton form factors, in particular
\begin{align}
\mathcal H(\xi)=\int_{-1}^{1}\frac{dx}{\xi} C(x/\xi) H(x,\xi)\,.
\label{CFF-H}
\end{align}
In order to estimate the size of the NNLO correction to the Compton form factor $\mathcal H(\xi)$ we use the GPD model
from Ref.~\cite[Eq.~(3.331)]{Belitsky:2005qn}, which is based
on the so-called double-distributions ansatz and allows for a simple analytic representation:
\begin{align}
H(x,\xi)&=\frac{(1-n/4)}{\xi^3}\biggl[\theta(x+\xi)\left(\frac{x+\xi}{1+\xi}\right)^{2-n}\Big(\xi^2-x+(2-n)\xi(1-x)\Big)
-(\xi\to -\xi)\biggr]\, .
\label{GPDmodel}
\end{align}
%
\begin{figure}[t]
\begin{center}
\includegraphics[width=0.62\linewidth]{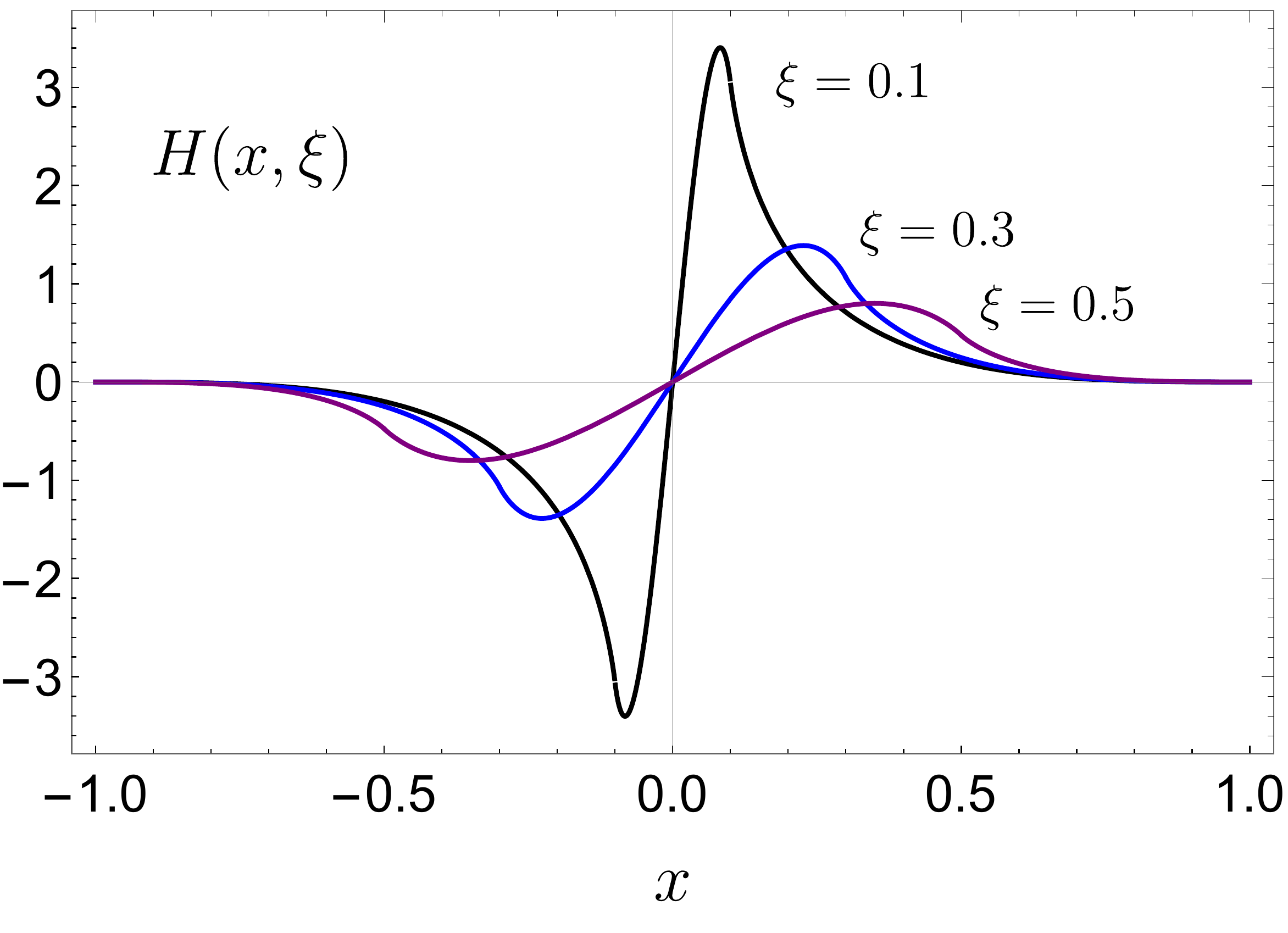}
\end{center}
\caption{The C-even part of the GPD model (taken from \cite[Eq.~(3.331)]{Belitsky:2005qn})
used for the calculation of the Compton form factor $\mathcal H$.
}
\label{fig:HGPD}
\end{figure}
(An overall normalization is irrelevant for our purposes so we omit it).
We use the value of the parameter $n=1/2$ which corresponds to a valence-like PDF $q(x)\sim x^{-1/2} (1-x)^3$ in the forward limit.
The $C$-even part of the GPD \eqref{GPDmodel}, $H(x,\xi)-H(-x,\xi)$, is shown in Fig.~\ref{fig:HGPD} for several
values of $\xi$.

\begin{figure}
\begin{center}
\includegraphics[width=0.47\linewidth]{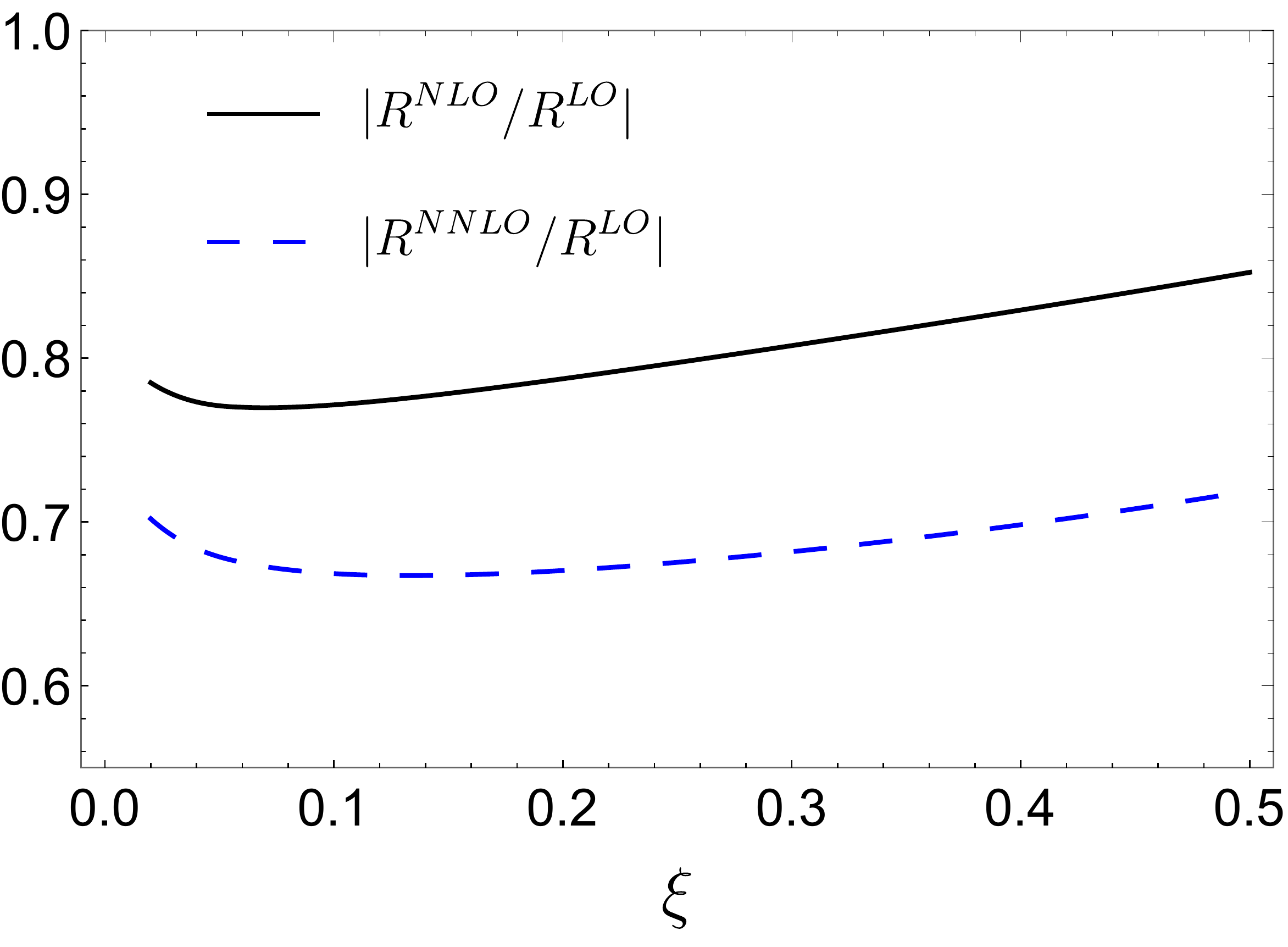}
\hskip 5mm
\includegraphics[width=0.47\linewidth]{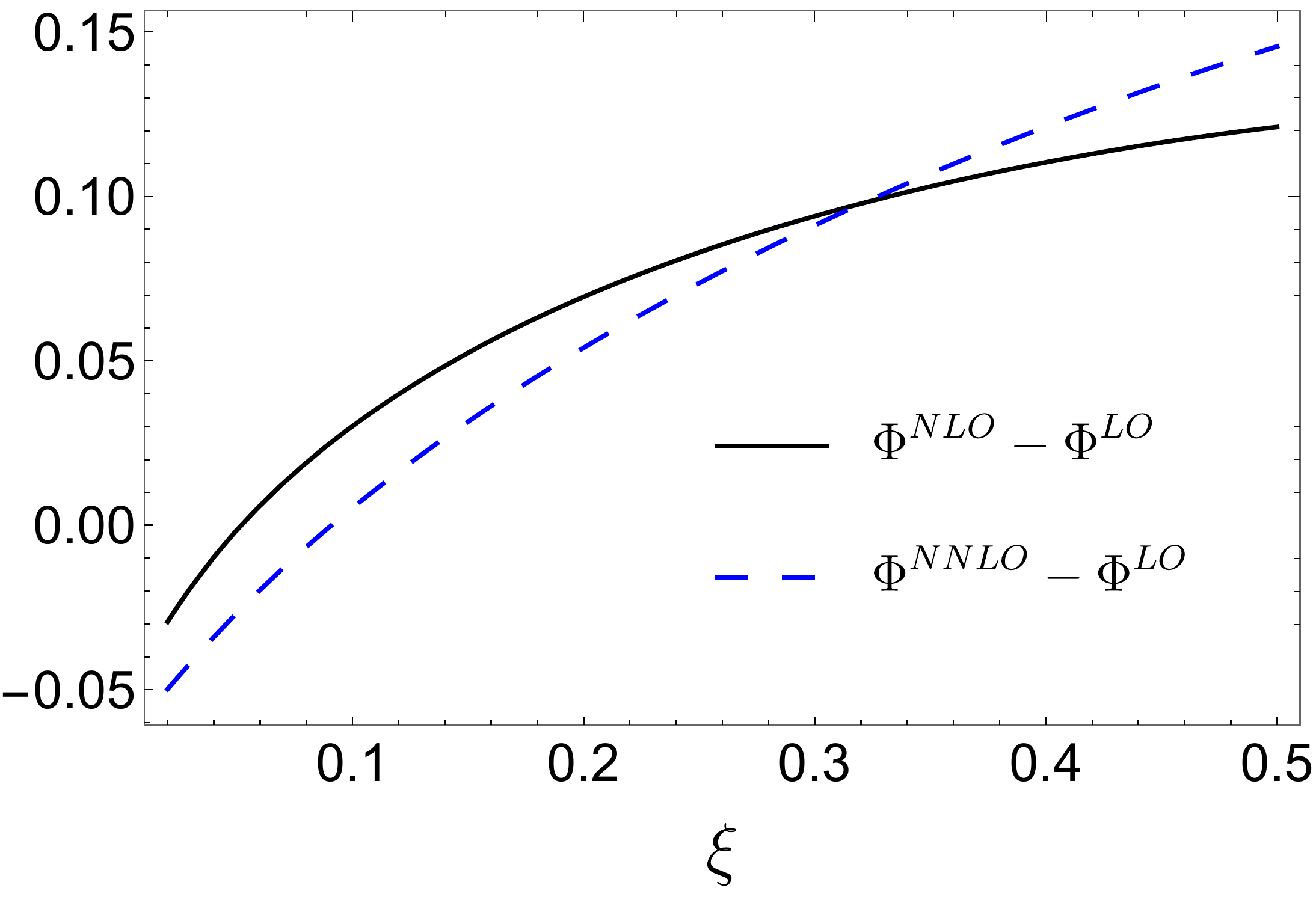}
\end{center}
\caption{
Higher-order QCD corrections to the Compton form factor $\mathcal H(\xi)$ for the GPD model in Eq.~\eqref{GPDmodel}.
The ratios of the Compton form factor calculated to the NNLO and NLO accuracy with respect to the tree-level
are shown for the absolute value and the phase of $\mathcal H(\xi)$, Eq.~\eqref{R-Phi}, on the left and the right panels,
respectively.
}
\label{fig:CFF-H}
\end{figure}

For a numerical evaluation of the contribution to the integral in Eq.~\eqref{CFF-H} from the DGLAP region
it proves to be convenient to shift the integration contour in the complex plane. We have checked that the
results do not depend on the shape of the integration contour, which is a good test of numerical accuracy.
The results are presented in Fig.\,\ref{fig:CFF-H}.
Following \cite{Kumericki:2007sa} we show the ratios for the absolute value and the phase of the Compton form factor,
\begin{align}
\mathcal H(\xi)= R(\xi)\, e^{i\Phi(\xi)}\,,
\label{R-Phi}
\end{align}
calculated to NNLO and NLO accuracy and normalized to the LO.
One sees that the NNLO correction to the absolute value of the Compton form factor $\mathcal H$ is quite large: it is only about factor two
smaller than the NLO correction and decreases the Compton form factor by about 10\% in the whole kinematic range.
The NNLO correction for the phase proves to be much smaller.

\section{Summary }\label{sec:summary}

Using an approach based on conformal symmetry~\cite{Braun:2013tva} we have calculated the two-loop CF in DVCS
in $\overline{\text{MS}}$ scheme
for the flavor-nonsinglet vector contributions.
Analytic expressions for the CF in momentum fraction space at $\mu=Q$ are
presented in Eqs.~\eqref{C2star}, \eqref{eq:d=4-result} and in Sect.~\ref{sec:scale} for an arbitrary scale.
Numerical estimates in Sect.~\ref{sec:numerics} suggest that the two-loop contribution gives rise
to a $\sim 10\%$ correction to the Compton form factor, which is significantly above the projected accuracy
at the  JLAB 12 GeV facility and the Electron Ion Collider.

We find an interesting hierarchy of different contributions to the two-loop CF, suggesting that the perturbative series in
conformal QCD at the critical coupling is converging much faster than in the physical case. It would be interesting to check
whether a similar hierarchy holds in other QCD examples, where the first few terms in perturbative expansion are known. The
corresponding study goes beyond the tasks of this work.

\section*{Acknowledgments}
\addcontentsline{toc}{section}{Acknowledgments}

This study was supported by Deutsche Forschungsgemeinschaft (DFG) through the Research Unit FOR 2926, ``Next Generation pQCD for
Hadron Structure: Preparing for the EIC'', project number 40824754, DFG grant $\text{MO~1801/1-3}$, and
RSF project No~19-11-00131.


\appendix
\addcontentsline{toc}{section}{Appendices}
\renewcommand{\theequation}{\Alph{section}.\arabic{equation}}
\renewcommand{\thesection}{{\Alph{section}}}
\renewcommand{\thetable}{\Alph{table}}
\setcounter{section}{0} \setcounter{table}{0}
\section*{Appendices}

\section{$SU(2)$-invariant kernels}\label{App:KernelList}

We collect here the invariant kernels and their eigenvalues used in Sect.~\ref{sec:two-loop}.
Let
\begin{align}
\mathrm M_n[\omega]=\int_0^1 d\alpha\int_0^{\bar\alpha} \!d\beta\,\omega(\tau) (1-\alpha-\beta)^{n-1}\,,
\qquad \tau =\frac{\alpha\beta}{\bar\alpha\bar\beta}\,,
\end{align}
where $n$ is even. One obtains
\begin{align}
\mathrm M_n[1] &=\frac1{n(n+1)}\,,  &  \mathrm M_n[\Li_2(\tau)]&=\frac{2S_{-2}(n)+\zeta_2}{n(n+1)}\,,
\notag\\
\mathrm M_n[-\ln\bar \tau] &= \frac{1}{n^2(n+1)^2}\,, &  \mathrm M_n\left[\frac{\bar\tau}{2\tau}\ln\bar\tau\right]&= S_3(n)-\zeta_3\,,
\notag\\
\mathrm M_n[-\ln\tau] & =\frac{2 S_1(n)}{n(n+1)}\,, & \mathrm M_n\left[-\bar\tau\ln\bar\tau\right] &= 2S_{-3}(n)
-4S_{1,-2}(n)-2\zeta_2 S_1(n)+\frac12\zeta_3\,,
\notag\\
\mathrm M_n[\bar\tau]&= 2S_{-2}(n)+\zeta_2\,,
\end{align}
and
\begin {align}
\mathrm M_n\left[\bar\tau\left(\Li_2(\tau)+\frac12\ln^2\bar\tau\right)\right] &=2S_{-4}(n)+\frac7{10}\zeta_2^2\,,
\notag\\
\mathrm M_n\left[\frac{\bar\tau}{4\tau}\left(\Li_2(\tau)+\frac12\ln^2\bar\tau\right)\right] &= S_{1,3}(n)-\frac12 S_4(n) -\zeta_3
S_1(n) +\frac3{10}\zeta_2^2\,.
\end{align}
%




\providecommand{\href}[2]{#2}\begingroup\raggedright\endgroup

\end{document}